\documentclass[twocolumn,tighten,twocolappendix]{aastex63}
\pdfoutput=1 
\usepackage{amsmath,amstext}
\usepackage[T1]{fontenc}
\usepackage{ae,aecompl}
\usepackage[utf8]{inputenc}
\usepackage{apjfonts} 
\usepackage[figure,figure*]{hypcap}
\usepackage{enumitem}
\usepackage{bm}
\usepackage{graphicx}
\usepackage{lineno}
\usepackage{tikz}

\definecolor{lmc}{HTML}{cc78bc}
\definecolor{mw}{HTML}{0173b2}
\definecolor{group}{HTML}{029e73}
\definecolor{lcluster}{HTML}{de8f05}

\input{hyperlink-year-only-natbib-patch}

\newcommand*{\http}[1]{\href{http://#1}{#1}}
\newcommand*{\https}[1]{\href{https://#1}{#1}}

\shorttitle{SIDM Concerto}
\shortauthors{Nadler et al.}

\begin{document}

\title{SIDM Concerto: Compilation and Data Release of Self-interacting Dark Matter Zoom-in Simulations}

\author[0000-0002-1182-3825]{Ethan O.~Nadler}
\affiliation{Department of Astronomy \& Astrophysics, University of California, San Diego, La Jolla, CA 92093, USA}

\author[0000-0003-1723-8691]{Demao Kong}
\affiliation{Department of Physics \& Astronomy, University of California, Riverside, CA 92521, USA}

\author[0000-0002-5421-3138]{Daneng Yang}
\affiliation{Purple Mountain Observatory, Chinese Academy of Sciences, Nanjing 210023, People's Republic of China}

\author[0000-0002-8421-8597]{Hai-Bo Yu}
\affiliation{Department of Physics \& Astronomy, University of California, Riverside, CA 92521, USA}

\correspondingauthor{Ethan~O.~Nadler}
\email{enadler@ucsd.edu}

\label{firstpage}

\begin{abstract}
We present SIDM Concerto: $14$ cosmological zoom-in simulations in cold dark matter (CDM) and self-interacting dark matter (SIDM) models based on the Symphony and Milky Way-est suites. SIDM Concerto includes one Large Magellanic Cloud-- (LMC-) mass system (host mass $\sim 10^{11}~M_{\mathrm{\odot}}$), two Milky Way (MW) analogs ($\sim 10^{12}~M_{\mathrm{\odot}}$), two group-mass hosts ($\sim 10^{13}~M_{\mathrm{\odot}}$), and one low-mass cluster ($\sim 10^{14}~M_{\mathrm{\odot}}$). Each host contains $\approx 2\times 10^7$ particles and is run in CDM and one or more strong, velocity-dependent SIDM models. Our analysis of SIDM (sub)halo populations over seven subhalo mass decades reveals that (1) the fraction of core-collapsed isolated halos and subhalos peaks at a maximum circular velocity corresponding to the transition of the SIDM cross section from a $v^{-4}$ to $v^0$ scaling; (2) SIDM subhalo mass functions are suppressed by $\approx 50\%$ relative to CDM in LMC, MW, and group-mass hosts but are consistent with CDM in the low-mass cluster host; (3) subhalos' inner density profile slopes, which are more diverse in SIDM than in CDM, are sensitive to both the amplitude and shape of the SIDM cross section. Our simulations provide a benchmark for testing SIDM predictions with astrophysical observations of field and satellite galaxies, strong lensing systems, and stellar streams. Data products are publicly available at \url{https://doi.org/10.5281/zenodo.14933624}.
\end{abstract}

\keywords{\href{http://astrothesaurus.org/uat/353}{Dark matter (353)}; 
\href{http://astrothesaurus.org/uat/1083}{$N$-body simulations (1083)};
\href{http://astrothesaurus.org/uat/1880}{Galaxy dark matter halos (1880)}}

\section{Introduction}\label{sec:intro}

Self-interacting dark matter (SIDM) can potentially alleviate small-scale structure anomalies that challenge the standard collisionless cold dark matter (CDM) paradigm (for reviews, see \citealt{Bullock170704256,Tulin170502358,Adhikari220710638}). There has recently been significant progress in determining the SIDM parameter space---that is, the cross section per dark matter (DM) mass $\sigma/m$ as a function of the scattering velocity $v$---that can resolve these anomalies. For example, it is now understood that successful SIDM models must yield a diverse distribution of DM halo density profiles on small scales, rather than uniformly producing central constant-density cores (e.g., \citealt{Correa200702958,Silverman220310104}). This can be achieved by a strong, velocity-dependent interaction cross section that produces core formation in high-mass halos and gravothermal core collapse in low-mass halos (\citealt{Shah230816342,Ando240316633}). At fixed mass, this outcome depends on halos' assembly histories as encoded in their concentrations, leading to a larger diversity in SIDM (sub)halo densities compared with CDM (e.g., \citealt{Nadler230601830,Yang221113768}).

This preferred SIDM parameter space has been determined using data that span a wide range of scales. On galaxy cluster scales ($v\sim 1000~\mathrm{km\ s}^{-1}$), halo shapes, mergers, and density profiles place upper limits on the SIDM cross section of $\sigma/m \lesssim 0.1$ to $1~\mathrm{cm^2}~\mathrm{g}^{-1}$ (e.g., \citealt{Rocha12083025,Harvey150307675,Kaplinghat150803339,Sagunski200612515,Andrade201206611}). On galactic scales ($v\sim 100~\mathrm{km\ s}^{-1}$), the rotation curves of low surface brightness and ultradiffuse galaxies yield $\sigma/m\gtrsim 3~\mathrm{cm}^2~\mathrm{g}^{-1}$~\citep{Ren180805695,Nadler230601830,Roberts240715005}. On subgalactic scales ($v\sim 10~\mathrm{km\ s}^{-1}$), satellite galaxies' diverse central densities prefer core collapse in at least some low-mass subhalos, such that $\sigma/m\gtrsim 5$ to $10~\mathrm{cm}^2~\mathrm{g}^{-1}$~\citep{Read180506934,Valli171103502,Sameie190407872,Silverman220310104}; meanwhile, diffuse satellites like Antlia 2 and Crater II favor large cores that are difficult to explain via tidal stripping in CDM \citep{Sameie200606681,Zhang240104985}. Other small-scale anomalies hint at the existence of extremely dense low-mass (sub)halos, including a strong-lensing substructure~\citep{Nadler230601830} and the GD-1 stellar stream perturber~\citep{Zhang240919493}, which could favor SIDM models that reach $\sigma/m \sim 100~\mathrm{cm^2}~\mathrm{g}^{-1}$ for $v\lesssim 10~\mathrm{km\ s}^{-1}$.

These results point toward a velocity-dependent SIDM cross section that impacts halo structure in qualitatively different ways at various relative scattering velocities and thus as a function of halo mass \citep{Kaplinghat150803339,Nadler230601830,Correa200702958,Correa240309186}. In particular, preferred velocity-dependent SIDM models generally predict cores in $\gtrsim 10^{11}~M_{\mathrm{\odot}}$ halos (although the predicted diversity can be large even at this scale; e.g., \citealt{Kong250106413}) and lead to core collapse in an increasing fraction of lower-mass halos down to $\sim 10^8~M_{\mathrm{\odot}}$ (e.g., \citealt{Shah230816342,Ando240316633}). Thus, to make further progress, it is important to understand halo and subhalo populations in the SIDM models of interest over a wide dynamic range.

Cosmological SIDM simulations are critical in order to achieve robust predictions over the full range of scales discussed above~(e.g., see \citealt{Banerjee220307049} for a review). This effort remains challenging despite rapid growth in the number and variety of such simulations (e.g., \citealt{Despali1811025699,Despali220412502,Despali250112439,Nadler200108754,Nadler210912120,Nadler230601830,Nadler241213065,Chua201008562,Turner201002924,Fischer220502243,Fischer231007750,Vargya210414069,Silverman220310104,O'Neil221016328,Yang221113768,Correa240309186,Ragagnin240401383,Leonard240113727}). In particular, simulations are often performed with different SIDM implementations and numerical settings (e.g., mass and spatial resolution) in varying cosmic environments (e.g., zoom-in vs.\ uniform-resolution boxes). These differences can complicate the interpretation of comparisons between literature results (e.g., \citealt{Meskhidze220306035}). Meanwhile, a growing body of work based on high-resolution controlled (i.e., noncosmological) simulations shows that convergence can be difficult to achieve in certain regions of SIDM parameter space and that robust SIDM (sub)halo modeling can require significantly higher resolution than expected based on CDM convergence tests \citep{Fischer240300739,Fischer250606269,Mace240201604,Palubski240212452}.

In this context, we present SIDM Concerto, a suite of $14$ cosmological DM--only zoom-in simulations of CDM and SIDM models. SIDM Concerto is based on the Symphony and Milky Way-est CDM zoom-in simulation suites, which span four decades of host halo mass at roughly fixed mass and spatial resolution with respect to each host~\citep{Nadler220902675,Buch240408043}. These suites were analyzed with a unified pipeline, including halo finding and merger tree algorithms, which can otherwise impact the interpretation of simulation results~\citep{Knebe11040949,Srisawat13073577}. SIDM Concerto inherits these advantages, which allows us to self-consistently simulate and analyze SIDM (sub)halo populations over an unprecedented dynamic range. All of our simulations are run at one resolution level higher than the fiducial Symphony and Milky Way-est suites, which allows us to capture core collapse in low-mass subhalos at each host scale and assess the impact of mass and spatial resolution on our simulation results. Furthermore, the SIDM models we simulate are motivated by the latest observations of satellite galaxies~\citep{Yang221113768}, strong gravitational lensing substructure~\citep{Nadler230601830}, and stellar stream perturbers~\citep{Zhang240919493}.

We run SIDM Concerto zoom-in simulations at the following four host halo--mass scales:
\begin{enumerate}
    \item Large Magellanic Cloud (LMC) mass ($\sim 10^{11}~M_{\mathrm{\odot}}$), with an eye toward LMC-associated satellite galaxies in the Milky Way (MW) and external LMC-mass galaxies (e.g., \citealt{Kallivayalil180501448,Carlin240917437});
    \item MW-mass ($\sim 10^{12}~M_{\mathrm{\odot}}$), with an eye toward satellite galaxies in the MW and in external MW-mass galaxies (e.g., \citealt{Drlica-Wagner191203302,Carlsten220300014,Mao240414498}), and stellar streams (e.g., \citealt{Bonaca240519410});
    \item Group mass ($\sim 10^{13}~M_{\mathrm{\odot}}$), with an eye toward strong lensing flux ratio statistics~\citep{Nierenberg190806344,Nierenberg230910101} and gravitational imaging~\citep{Vegetti09100760,Vegetti12013643,Hezaveh160101388};
    \item Low-mass galaxy cluster (L-Cluster; $\sim 10^{14}~M_{\mathrm{\odot}}$), with an eye toward strong~\citep{Meneghetti200904471} and weak~\citep{Banerjee190612026,Bhattacharyya210608292,Adhikari240105788} gravitational lensing data.
\end{enumerate}
At each host-mass scale, we simulate CDM and one or more strong, velocity-dependent SIDM models that are preferred by small-scale structure data and compatible with cluster constraints. Each host is resolved with $\approx 2\times 10^7$ particles, which allows us to study (sub)halo abundances and density profiles down to $\approx 10^{-5}$ and $10^{-4}$ times each host mass, respectively. SIDM Concerto includes the zoom-ins presented in \cite{Yang221113768} and \cite{Nadler241213065} at the MW scale and in \cite{Nadler230601830} at the group scale, along with new LMC, Group, and L-Cluster zoom-ins. Our work is accompanied by a public data release of halo catalogs, merger trees, and particle snapshots.

This paper is organized as follows. Section~\ref{sec:model} describes the SIDM models we simulate. Section~\ref{sec:simulations} provides technical details about our simulations and describes our analysis procedures. Section~\ref{sec:results} presents population statistics of isolated halos and subhalos from our simulations. Section~\ref{sec:density} presents subhalo density profiles. Section~\ref{sec:discussion} provides a discussion of our results, and Section~\ref{sec:conclusions} concludes.

We adopt cosmological parameters that match the corresponding Symphony suites: $h = 0.7$, $\Omega_{m} = 0.286$, $\Omega_{\Lambda} = 0.714$, $\sigma_8 = 0.82$, and $n_s=0.96$ for the LMC, MW, and Group hosts \citep{Hinshaw_2013}, and $h = 0.7$, $\Omega_{m} = 0.3$, $\Omega_{\Lambda} = 0.7$, $\sigma_8 = 0.85$, and $n_s=0.96$ for the L-Cluster hosts \citep{Banerjee190612026,Bhattacharyya210608292}. Virial masses are defined using the \cite{Bryan_1998} overdensity. Throughout, ``log'' refers to the base-10 logarithm.


\section{SIDM Models}
\label{sec:model}

We consider SIDM models with a differential scattering cross section~\citep{Ibe:2009mk,Yang220503392}
\begin{equation}
    \frac{\mathrm{d}\sigma}{\mathrm{d}\cos\theta} = \frac{\sigma_0w^4}{2\left[w^2+v^2\sin^2(\theta/2)\right]^2},\label{eq:xsec}
\end{equation}
where $v$ and $\theta$ are the relative scattering velocity and angle, respectively, $\sigma_0$ sets the cross-section amplitude, and $w$ sets the velocity at which the cross section transitions from a $v^{-4}$ to $v^{0}$ velocity scaling. We simulate three SIDM models in this work:
\begin{enumerate}
    \item \emph{GroupSIDM}: $\sigma_0/m = 147.1~\mathrm{cm}^2~\mathrm{g}^{-1}$, $w=120~\mathrm{km\ s}^{-1}$. \cite{Nadler230601830} implemented this model in a simulation of a Group-mass host (Halo352), which is part of our suite. In that study, the GroupSIDM cross section was shown to produce both extremely high-concentration core-collapsed subhalos of the Group host, analogous to an observed lensing perturber, and extremely low-concentration core-forming isolated halos, analogous to observed ultradiffuse galaxies~\citep{Kong220405981,Mancera240406537}.
    \item \emph{GroupSIDM-70}: $\sigma_0/m = 70~\mathrm{cm}^2~\mathrm{g}^{-1}$, $w=120~\mathrm{km\ s}^{-1}$. We introduce this model to study the impact of varying $\sigma_0/m$ at fixed $w$ (with respect to GroupSIDM).
    \item \emph{MilkyWaySIDM}: $\sigma_0/m = 147.1~\mathrm{cm}^2~\mathrm{g}^{-1}$, $w=24.3~\mathrm{km\ s}^{-1}$. This model was introduced in \cite{Yang221113768} and is similar to those in \cite{Correa200702958} and \cite{Turner201002924}. \cite{Yang221113768} showed that this model diversifies (sub)halo density profiles using an MW-mass zoom-in (Halo416) that is part of our suite. It allows us to study the impact of varying $w$ at fixed $\sigma_0/m$ (with respect to GroupSIDM).
\end{enumerate}

Figure~\ref{fig:xsec} shows the corresponding effective cross sections, host halo velocity scales, and subhalo velocity scales as a function of maximum circular velocity $V_{\mathrm{max}}$. Effective cross sections are calculated following \cite{Yang220503392}, \cite{Yang221113768}, and \cite{Nadler230601830}. Our simulated subhalo populations span $V_{\mathrm{max}}$ values that probe different cross sections depending on the host-mass scale and SIDM model, represented by the vertical shaded bands in Figure~\ref{fig:xsec}. For example, in the GroupSIDM and GroupSIDM-70 models, most subhalos we resolve probe the high-amplitude $v^0$ scattering regime. The host halos set other relevant velocity scales, shown by the dashed vertical lines in Figure~\ref{fig:xsec}. Specifically, subhalo--host halo interactions at these velocity scales can lead to evaporation and accelerate tidal disruption (e.g., \citealt{Nadler200108754,Zeng211000259}). Most of the host halo--SIDM model combinations we consider correspond to moderately strong subhalo--host halo interactions ($\gtrsim 1~\mathrm{cm}^2~\mathrm{g}^{-1}$).


\section{Simulations and Analyses}
\label{sec:simulations}

\subsection{Zoom-in Simulations}
\label{sec:zoom-ins}

We run all simulations using modified versions of \textsc{Gadget-2} \citep{Springel0505010} that include DM self-interactions \citep{Yang220503392,Nadler230601830,Yang221113768}. For the MilkyWaySIDM simulations, our implementation captures both the velocity and angular dependence of the differential SIDM cross section (Equation~\ref{eq:xsec}). In particular, scattering probabilities are calculated using both the relative velocity and angle of each simulation particle pair. For the GroupSIDM and GroupSIDM-70 simulations, we instead implement the velocity-dependent but isotropic viscosity cross section, such that scattering probabilities only depend on the relative velocity of each simulation particle pair. This isotropic viscosity cross-section implementation has been shown to match the results of simulations that use the full, angularly dependent SIDM scattering cross section \citep{Yang220503392}.

In all cases, we generate initial conditions using \textsc{MUSIC}~\citep{Hahn11036031}, with high-resolution regions that span $\sim 10\times$ the virial radius $R_{\mathrm{vir}}$ of each host. The numerical settings for each host are described below; we present convergence tests using lower-resolution resimulations of every host in Appendix~\ref{sec:convergence}. Figure~\ref{fig:vis_main} visualizes several of our hosts, and Table~\ref{tab:summary} summarizes key properties of each simulation.

\subsubsection{LMC Mass}

We resimulate Halo104 from the Symphony LMC suite in CDM and GroupSIDM with six nested \textsc{MUSIC} refinement regions centered on the LMC-mass host. The highest-resolution particle mass is $m_{\mathrm{part}}=6.3\times 10^3~M_{\mathrm{\odot}}$ and the comoving gravitational softening is $\epsilon=40~\mathrm{pc}~h^{-1}$. These resimulations are presented here for the first time.

\begin{figure}[t!]
\hspace{-4mm}
\includegraphics[trim={0 0.5cm 0 0cm},width=0.5\textwidth]{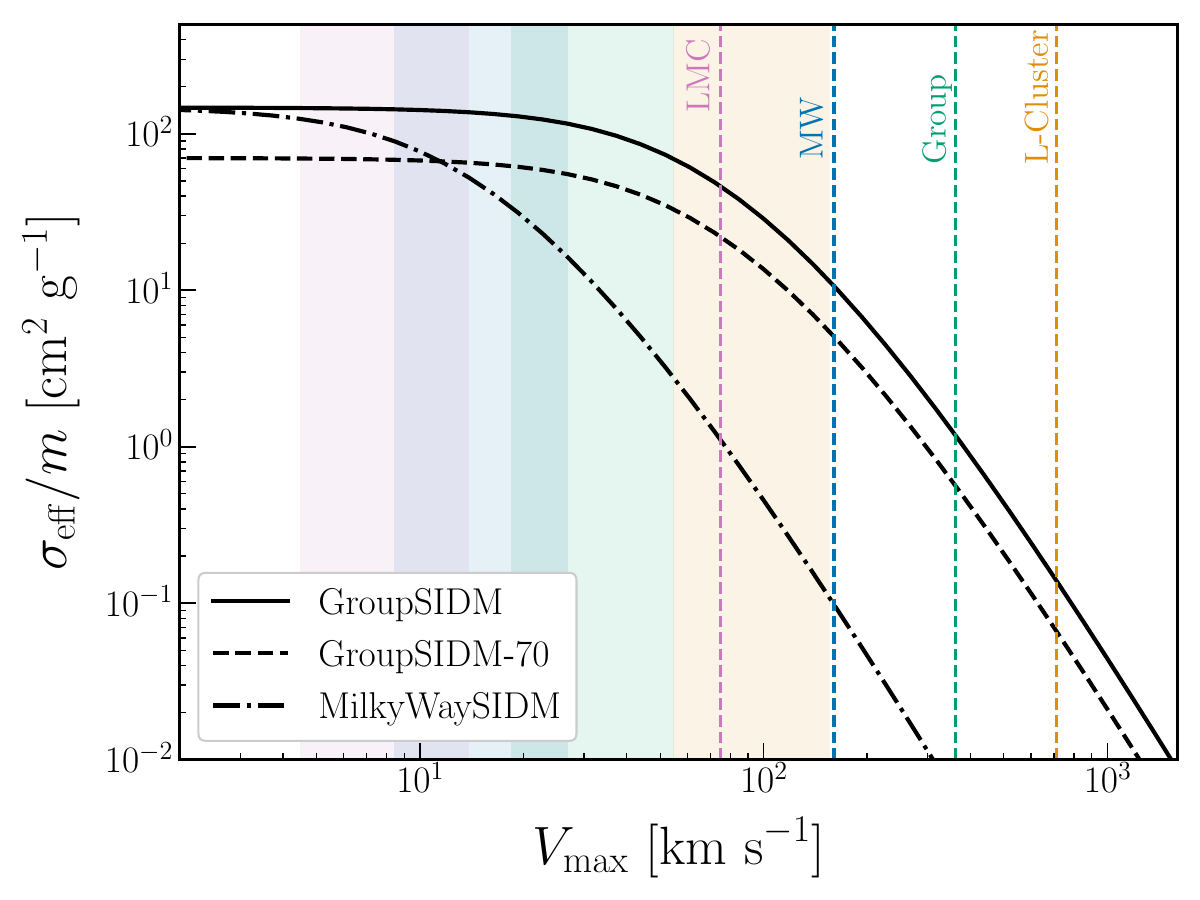}
   \caption{Effective cross sections for our GroupSIDM (solid black), GroupSIDM-70 (dashed black), and MilkyWaySIDM (dotted--dashed black) SIDM models. Vertical shaded bands show the $1\sigma$ range of $V_{\mathrm{max}}$ for CDM subhalos with $M_{\mathrm{vir}}>2000m_{\mathrm{part}}$, and vertical dashed lines show corresponding CDM host halos' average $V_{\mathrm{max}}$ values.}
    \label{fig:xsec}
\end{figure}

\begin{figure*}[t!]
\centering
\vspace{0.5cm}
\includegraphics[trim={0 0cm 0 0cm},width=0.8\textwidth]{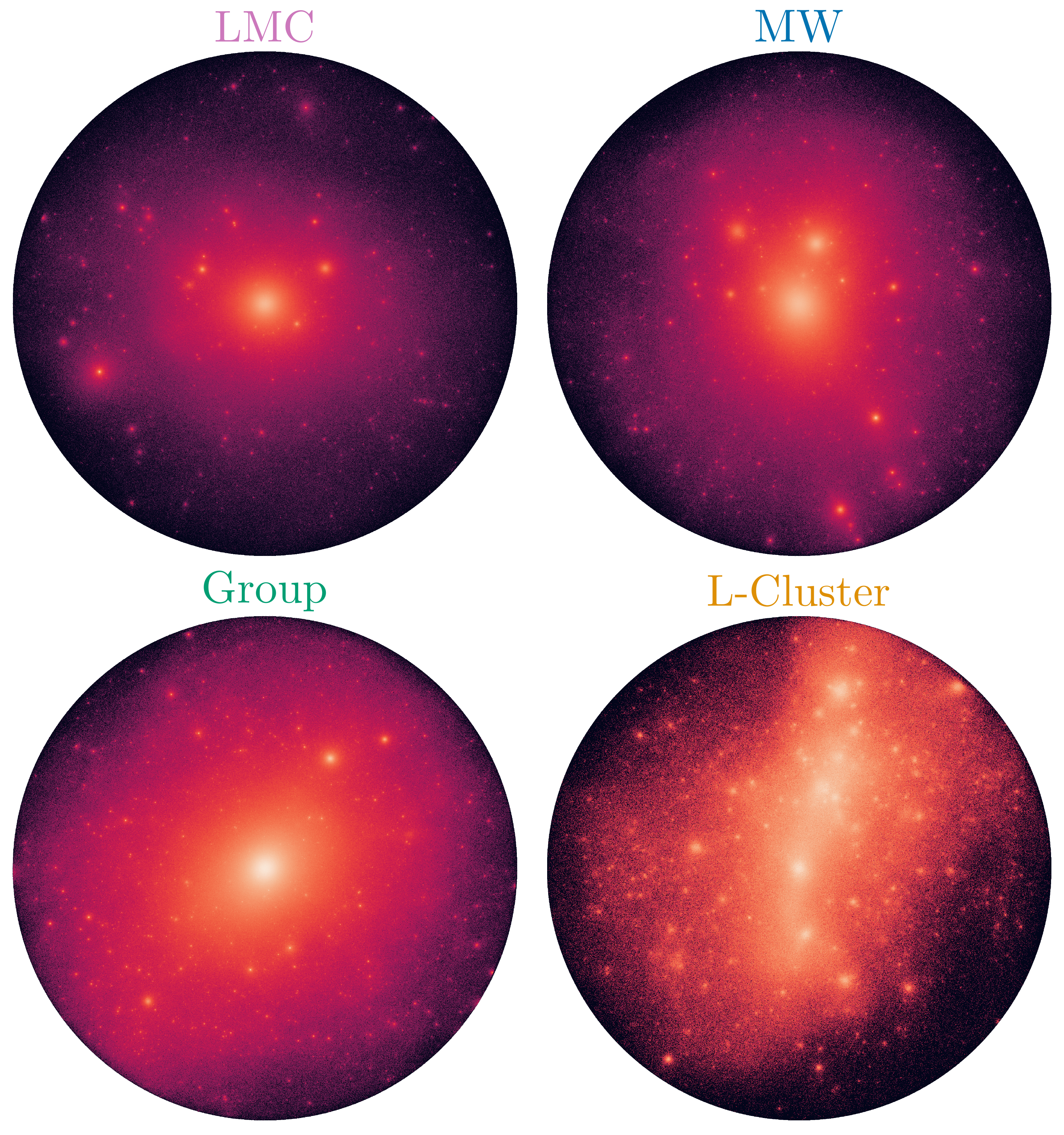}
    \caption{Projected DM density maps for GroupSIDM simulations of LMC Halo104 (top left), MW Halo004 (top right), Group Halo352 (bottom left), and L-Cluster Halo000 (bottom right). Each visualization spans the virial radius of each host and is created using \textsc{meshoid}.}
    \label{fig:vis_main}
\end{figure*}

\begin{deluxetable*}{{cccccccc}}[t!]
\centering
\tablecolumns{8}
\tablecaption{Summary of SIDM Concerto Simulations.}
\tablehead{\colhead{Suite} & \colhead{Host} & \colhead{Model} & \colhead{$\log(M_{\mathrm{host}}/M_{\mathrm{\odot}})$} & \colhead{$R_{\mathrm{vir}}~[\mathrm{kpc}]$} & \colhead{$m_{\mathrm{part}}~[M_{\mathrm{\odot}}]$} & \colhead{$\epsilon~[\mathrm{pc}~h^{-1}]$} & \colhead{Color and Linestyle}}
\startdata 
\hline \hline
LMC & 
104 &
CDM &
10.9 &
117.3 &
$6.3\times 10^3$ & 
40 &
\begin{tikzpicture}[yscale=0.5] \draw [line width=0.45mm,dotted,black] (0,-1) -- (1,-1) node[right]{};; \end{tikzpicture} \\ 
LMC & 
104 & 
GroupSIDM & 
10.9 &
115.5 &
$6.3\times 10^3$ & 
40 &
\begin{tikzpicture}[yscale=0.5] \draw [line width=0.25mm,lmc] (0,-1) -- (1,-1) node[right]{};; \end{tikzpicture}
\\
\hline \hline
MW & 
004 &
CDM &
12.0 &
263.1 &
$5.0\times 10^4$ & 
80 &
\begin{tikzpicture}[yscale=0.5] \draw [line width=0.45mm,dotted,black] (0,-1) -- (1,-1) node[right]{};; \end{tikzpicture} \\ 
MW & 
004 & 
GroupSIDM & 
12.0 &
261.6 &
$5.0\times 10^4$ & 
80 &
\begin{tikzpicture}[yscale=0.5] \draw [line width=0.25mm,mw] (0,-1) -- (1,-1) node[right]{};; \end{tikzpicture}
\\
MW & 
004 & 
MilkyWaySIDM & 
12.0 &
263.0 &
$5.0\times 10^4$ & 
80 &
\begin{tikzpicture}[yscale=0.5] \draw [line width=0.35mm,dashed,mw] (0,-1) -- (1,-1) node[right]{};; \end{tikzpicture}
\\
MW & 
416 &
CDM &
12.2 &
307.2 &
$5.0\times 10^4$ & 
80 &
\begin{tikzpicture}[yscale=0.5] \draw [line width=0.45mm,dotted,black] (0,-1) -- (1,-1) node[right]{};; \end{tikzpicture} \\ 
MW & 
416 & 
MilkyWaySIDM & 
12.2 &
307.2 &
$5.0\times 10^4$ & 
80 &
\begin{tikzpicture}[yscale=0.5] \draw [line width=0.35mm,dashed,mw] (0,-1) -- (1,-1) node[right]{};; \end{tikzpicture}
\\
\hline \hline
Group & 
352 &
CDM &
13.1 &
609.5 &
$4.0\times 10^5$ & 
170 &
\begin{tikzpicture}[yscale=0.5] \draw [line width=0.45mm,dotted,black] (0,-1) -- (1,-1) node[right]{};; \end{tikzpicture} \\ 
Group & 
352 & 
GroupSIDM & 
13.1 &
608.3 &
$4.0\times 10^5$ & 
170 &
\begin{tikzpicture}[yscale=0.5] \draw [line width=0.25mm,group] (0,-1) -- (1,-1) node[right]{};; \end{tikzpicture}
\\
Group & 
352 & 
GroupSIDM-70 & 
13.1 &
608.7 &
$4.0\times 10^5$ & 
170 &
\begin{tikzpicture}[yscale=0.5] \draw [line width=0.35mm,dash dot,group] (0,-1) -- (1,-1) node[right]{};; \end{tikzpicture}
\\
Group & 
962 &
CDM &
13.5 &
818.4 &
$4.0\times 10^5$ & 
170 &
\begin{tikzpicture}[yscale=0.5] \draw [line width=0.45mm,dotted,black] (0,-1) -- (1,-1) node[right]{};; \end{tikzpicture} \\ 
Group & 
962 & 
GroupSIDM-70 & 
13.5 &
815.9 &
$4.0\times 10^5$ & 
170 &
\begin{tikzpicture}[yscale=0.5] \draw [line width=0.35mm,dash dot,group] (0,-1) -- (1,-1) node[right]{};; \end{tikzpicture}
\\
\hline \hline
L-Cluster & 
000 &
CDM &
14.2 &
1379.5 &
$2.7\times 10^7$ & 
600 &
\begin{tikzpicture}[yscale=0.5] \draw [line width=0.45mm,dotted,black] (0,-1) -- (1,-1) node[right]{};; \end{tikzpicture} \\ 
L-Cluster & 
000 & 
GroupSIDM & 
14.2 &
1382.1 &
$2.7\times 10^7$ & 
600 &
\begin{tikzpicture}[yscale=0.5] \draw [line width=0.25mm,lcluster] (0,-1) -- (1,-1) node[right]{};; \end{tikzpicture}
\\
\hline \hline
\enddata
{\footnotesize \tablecomments{The first two columns respectively list the host halo suite and name. The third column lists the DM model. The fourth and fifth columns list the host halo virial mass and virial radius at $z=0$, and the sixth and seventh columns list the high-resolution particle mass and comoving gravitational softening. The last column lists the color and line style.}}
\label{tab:summary}
\end{deluxetable*}

\subsubsection{MW Mass}

We include the CDM and MilkyWaySIDM versions of Halo416 presented in \cite{Yang221113768}. These simulations use five nested \textsc{MUSIC} refinement regions centered on the MW-mass host, with $m_{\mathrm{part}}=5\times 10^4~M_{\mathrm{\odot}}$ and $\epsilon=80~\mathrm{pc}~h^{-1}$ in the highest-resolution region. This host was originally simulated in \cite{Mao150302637} and was then included in the Symphony MW-mass suite~\citep{Nadler220902675}. This host contains a realistic LMC analog and experiences a Gaia--Sausage--Enceladus (GSE)-like major merger at $z\approx 2$. Note that we use the version of this host with $n_s=1$ as in \cite{Mao150302637}, rather than $n_s=0.96$ as in \cite{Nadler220902675}, because this setting results in a more realistic LMC infall time and a present-day LMC distance of $\approx 50~\mathrm{kpc}$.

We also include the CDM and MilkyWaySIDM versions of Halo004 from COZMIC III \citep{Nadler241213065}, and we resimulate this host in GroupSIDM. Numerical settings and cosmological parameters match Halo416, except that we use $n_s=0.96$ rather than $n_s=1$. Note that Halo004 contains an LMC analog and experiences an early major merger with a GSE analog. This host was originally presented in the Milky Way-est suite~\citep{Buch240408043}.

\subsubsection{Group Mass}

We include the CDM and GroupSIDM versions of Halo352 from \cite{Nadler230601830}; we also resimulate this host in GroupSIDM-70. These simulations use four nested \textsc{MUSIC} refinement regions centered on the Group-mass host, with $m_{\mathrm{part}}=4\times 10^5~M_{\mathrm{\odot}}$ and $\epsilon=170~\mathrm{pc}~h^{-1}$ in the highest-resolution region. This host was originally part of the Symphony Group-mass suite~\citep{Nadler220902675}.

We also resimulate a higher-mass Group host from Symphony, Halo962, in CDM and GroupSIDM-70. Numerical settings and cosmological parameters for this host are the same as for Halo352.

\subsubsection{L-Cluster}

We resimulate Halo000 from the L-Cluster suite in CDM and GroupSIDM using five nested \textsc{MUSIC} refinement regions centered on the L-Cluster host, with $m_{\mathrm{part}}=2.7\times 10^7~M_{\mathrm{\odot}}$ and $\epsilon=600~\mathrm{pc}~h^{-1}$ in the highest-resolution region. This host was originally presented in \cite{Bhattacharyya210608292} and was included in the Symphony L-Cluster suite~\citep{Nadler220902675}. \cite{Bhattacharyya210608292} presented a lower-resolution version of this host in a velocity-dependent SIDM model with $\sigma_0/m\sim 1~\mathrm{cm}^2~g^{-1}$, which is smaller than any of the models considered in this work.

\begin{figure*}[t!]
\centering
\includegraphics[width=\textwidth]{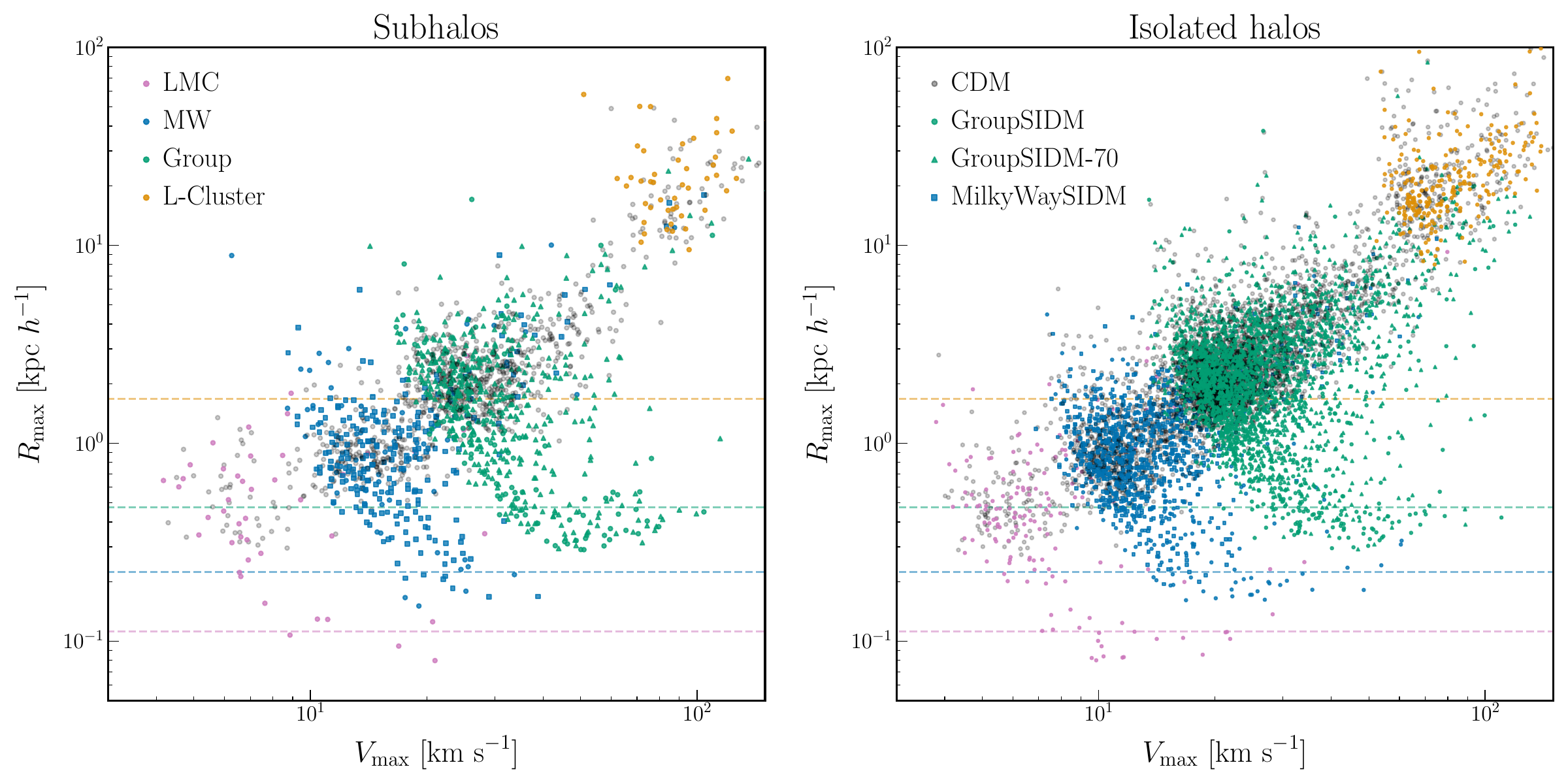}
    \caption{$R_{\mathrm{max}}$--$V_{\mathrm{max}}$ relation for subhalos (left panel) and isolated halos (right panel) in CDM (gray circles), GroupSIDM (colored circles), GroupSIDM-70 (colored triangles), and MilkyWaySIDM (colored squares). In both panels, we restrict to (sub)halos with $M_{\mathrm{vir}}>2000 m_{\mathrm{part}}$. Dashed horizontal lines show the approximate spatial resolution scale, $2.8 \epsilon$, for each suite. For hosts simulated with multiple SIDM cross sections, we combine all SIDM results in these figures. Note that GroupSIDM-70 is only simulated at the Group scale, while MilkyWaySIDM is only simulated at the MW scale.}
    \label{fig:rmax_vmax}
\end{figure*}

\subsection{(Sub)halo Catalogs and Merger Trees}
\label{sec:analysis}

For our main analyses, we generate (sub)halo catalogs and merger trees using {\sc Rockstar} and {\sc consistent-trees} (hereafter \textsc{RCT}; \citealt{Behroozi11104372,Behroozi11104370}). In Appendix~\ref{sec:symfind}, we compare our main results to those derived using the particle-tracking subhalo finder \textsc{Symfind}~\citep{Mansfield230810926}.

In each simulation, we analyze all subhalos above that are above the mass thresholds listed below and within the virial radius of the main host at $z=0$. In some sections, we also analyze isolated halos above the relevant mass threshold within $10~R_{\mathrm{vir}}$ of each host, corresponding to $\sim 1~\mathrm{Mpc}$, $3~\mathrm{Mpc}$, $6~\mathrm{Mpc}$, and $10~\mathrm{Mpc}$ for the LMC, MW, Group, and L-Cluster hosts, respectively. The fraction of low-resolution particles within these regions is negligible in all cases.

Throughout, we only analyze the \emph{abundance} of (sub)halos with virial masses $M_{\mathrm{vir}}(z=0)>300 m_{\mathrm{part}}$, that is, $M_{\mathrm{vir}}>[1.9\times 10^6,1.5\times 10^7,1.2\times 10^8,8.1\times 10^9]~M_{\mathrm{\odot}}$ for the LMC, MW, Group, and L-Cluster hosts, respectively. Meanwhile, we only analyze the \emph{internal properties} (i.e., $V_{\mathrm{max}}$, $R_{\mathrm{max}}$, and density profiles) for (sub)halos with $M_{\mathrm{vir}}(z=0)>2000 m_{\mathrm{part}}$, that is, $M_{\mathrm{vir}}>[1.3\times 10^7,10^8,8\times 10^8,5.4\times 10^{10}]~M_{\mathrm{\odot}}$.\footnote{For each (sub)halo, $V_{\mathrm{max}}$ is defined as the maximum of $\sqrt{GM(<r)/r}$ where $M(<r)$ is the enclosed mass profile as a function of radius, and $R_{\mathrm{max}}$ is defined as the radius within each (sub)halo at which $V_{\mathrm{max}}$ occurs.}

When analyzing density profiles, we treat the spline softening length of $2.8\epsilon$ as the spatial resolution scale. Note that the ratio of the Plummer-equivalent gravitational softening $\epsilon$ to the mean interparticle spacing $\ell$ in all of our simulations is $\epsilon/\ell \approx 0.01$~\citep{Nadler220902675}, which implies that $2.8\epsilon \approx 0.028\ell$. This is a factor of $\approx 2$ smaller than the two-body relaxation `convergence radius' of $0.055\ell$ from \cite{Ludlow2019}, in line with their recommendations for CDM halos. Note that our simulations have sufficiently fine time stepping for these arguments to apply: \texttt{ErrTolIntAcc}$=0.01$ for the LMC, MW, and Group hosts and $0.05$ for the L-Cluster host. We demonstrate convergence of SIDM $V_{\mathrm{max}}$ and $R_{\mathrm{max}}$ distributions in Appendix~\ref{sec:convergence}.

\subsection{Parametric SIDM Model}
\label{sec:parametric}

We model gravothermal evolution using the parametric model for SIDM \citep{Yang230516176,Yang240610753}. This model predicts SIDM halos' density profile evolution using $V_{\mathrm{max}}$ and $R_{\mathrm{max}}$ evolution histories from matched CDM halos. It has been validated using $V_{\mathrm{max}}$, $R_{\mathrm{max}}$, and density profile evolution histories from controlled and cosmological simulations, including a subset of the MW (Halo416) and Group (Halo352) hosts presented here \citep{Yang240610753}.

We parameterize each (sub)halo's gravothermal evolution by $\tau\equiv t/t_c$, where $t$ is the time elapsed since halo formation and $t_c$ is the core-collapse timescale, which we discuss further in Section~\ref{sec:parametric}. Specifically, for each (sub)halo under consideration, we calculate
\begin{equation}
    \tau_0 = \int_{t_f}^{t_0} \frac{\mathrm{d}t}{t_c(t)},\label{eq:tau_0}
\end{equation}
where $t_0=13.6~\mathrm{Gyr}$ and $t_f$ is the formation time. Note that the SIDM model dependence is encoded in $t_c(t)$.

For each SIDM Concerto host, we apply the parametric model to the CDM simulation to predict $\tau_0$ for SIDM (sub)halos under a given cross-section model. Following \cite{Nadler241213065}, we define (sub)halos with $\tau_0<0.15$ as ``core-forming'' and (sub)halos with $\tau_0>0.75$ as ``core-collapsed.'' The former value selects the phase during which SIDM halos' central densities decrease, and the latter selects the phase when central densities exceed CDM \citep{Outmezguine220406568,Yang220502957}. We set the maximum value of $\tau_0$ to $1.1$, up to which the parametric model has been calibrated using controlled simulations \citep{Yang240610753}.

\section{Halo and Subhalo Population Statistics}
\label{sec:results}

We now study the $R_{\mathrm{max}}$--$V_{\mathrm{max}}$ relation (Section~\ref{sec:rmax_vmax}), gravothermal evolution timescales (Section~\ref{sec:tau_0}), and mass functions (Section~\ref{sec:shmf}) of (sub)halos in our simulations.

\subsection{$R_{\mathrm{max}}$--$V_{\mathrm{max}}$ Relations}
\label{sec:rmax_vmax}

The $R_{\mathrm{max}}$--$V_{\mathrm{max}}$ relation probes SIDM halos' gravothermal evolution \citep{Yang221113768,Yang230516176,Ando240316633}. In CDM, this relation is determined by the underlying mass--concentration relation; this correspondence is exact for Navarro--Frenk--White (NFW; \citealt{Navarro1997}) density profiles. While SIDM halos are generally not well described by NFW profiles, shifts in the $R_{\mathrm{max}}$--$V_{\mathrm{max}}$ relation relative to CDM can indicate core formation (which shifts (sub)halos toward smaller $V_{\mathrm{max}}$ and larger $R_{\mathrm{max}}$) and collapse (which shifts (sub)halos toward larger $V_{\mathrm{max}}$ and smaller $R_{\mathrm{max}}$).

\begin{figure*}[t!]
\centering
\includegraphics[trim={0 0.5cm 0 0cm},width=\textwidth]{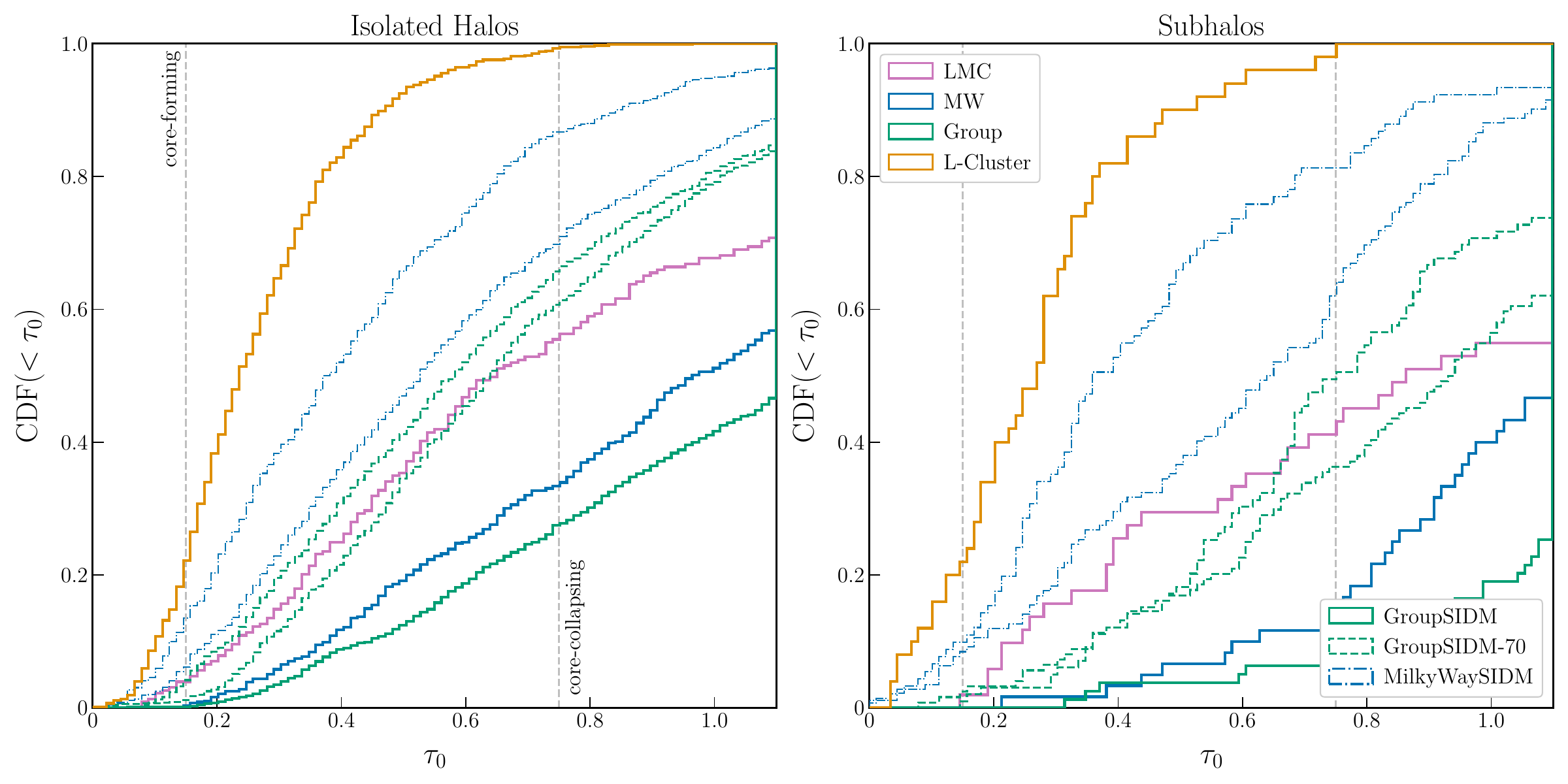}
    \caption{Normalized cumulative distributions of the gravothermal evolution timescale, $\tau_0$, calculated using the parametric SIDM model from \cite{Yang230516176,Yang240610753} for isolated halos (left panel) and subhalos (right panel) with $M_{\mathrm{vir}}>2000m_{\mathrm{part}}$. Results are shown for each SIDM Concerto host halo-mass scale (different colors) and for various SIDM models (different line styles). (Sub)halos with $\tau_0<0.15$ have decreasing central densities; (sub)halos with $\tau_0>0.75$ have central densities that exceed their CDM counterparts.}
    \label{fig:tau_0}
\end{figure*}

Figure~\ref{fig:rmax_vmax} shows the $R_{\mathrm{max}}$--$V_{\mathrm{max}}$ relation for subhalos (left panel) and isolated halos (right panel) in all SIDM Concerto simulations. Relative to CDM (gray points), SIDM (sub)halos (colored points) display larger scatter in the $R_{\mathrm{max}}$--$V_{\mathrm{max}}$ relation at all host-mass scales. In the L-Cluster host, the SIDM relation is shifted in the core-forming direction, while in the lower-mass hosts, the shift is mainly in the core-collapse direction. This is consistent with our expectation from Figure~\ref{fig:xsec}, since the L-Cluster (sub)halos probe a lower-amplitude part of the SIDM cross section than the subhalos of the lower-mass hosts. Deeply collapsed (sub)halos have values of $R_{\mathrm{max}}$ comparable to the convergence radius at each host mass, shown by the dashed horizontal lines in Figure~\ref{fig:rmax_vmax}. The fact that we cannot resolve (sub)halos with $R_{\mathrm{max}}$ values well below these radii is a consequence of our $M_{\mathrm{vir}}$ cut; (sub)halos with significantly smaller values of $R_{\mathrm{max}}$ than $2.8\epsilon$ would not be well resolved.

Comparing the left and right panels of Figure~\ref{fig:rmax_vmax}, we find a larger abundance of cored (high-$R_{\mathrm{max}}$) isolated halos compared to subhalos at each host-mass scale. There are several contributing factors: (1) subhalos are more concentrated than isolated halos~\citep{Moline160304057}, which accelerates their gravothermal evolution~\citep{Essig180901144,Yang221113768}; (2) subhalos with prominent cores are more susceptible to tidal disruption~\citep{Errani221001131}; (3) tidal stripping accelerates core collapse~\citep{Nishikawa190100499,Sameie190407872}. The relatively high abundance of core-collapsed subhalos (along with the lack of surviving cored subhalos) is most pronounced in the Group simulations, consistent with the results from \cite{Nadler230601830}.

\subsection{Gravothermal Evolution Timescales}
\label{sec:tau_0}

Figure~\ref{fig:tau_0} shows the cumulative distribution of $\tau_0$ (Equation~\ref{eq:tau_0}) for isolated halos (left panel) and subhalos (right panel) in each SIDM Concerto simulation.\footnote{The cumulative distributions in Figure~\ref{fig:tau_0} jump to unity at $\tau_0=1.1$ because we set the maximum value of $\tau_0$ from our parametric model predictions to $1.1$, even if the model predicts larger values of $\tau_0$, according to Section~\ref{sec:parametric}.} Consistent with the $R_{\mathrm{max}}$--$V_{\mathrm{max}}$ relation, nearly no (sub)halos in the L-Cluster suite have $\tau_0>0.75$, which implies that their central densities do not exceed their CDM counterparts; we confirm this by measuring subhalo density profiles in Section~\ref{sec:density}. The core-collapsed fraction is generally higher in the lower-mass suites, although there is significant variation between SIDM models at each host-mass scale. In the GroupSIDM model, we find that the core-collapsed fraction peaks at the Group host-mass scale and significantly decreases in lower-mass hosts; we interpret this result below.

\begin{figure}[t!]
\hspace{-4mm}
\includegraphics[trim={0 0.5cm 0 0cm},width=0.5\textwidth]{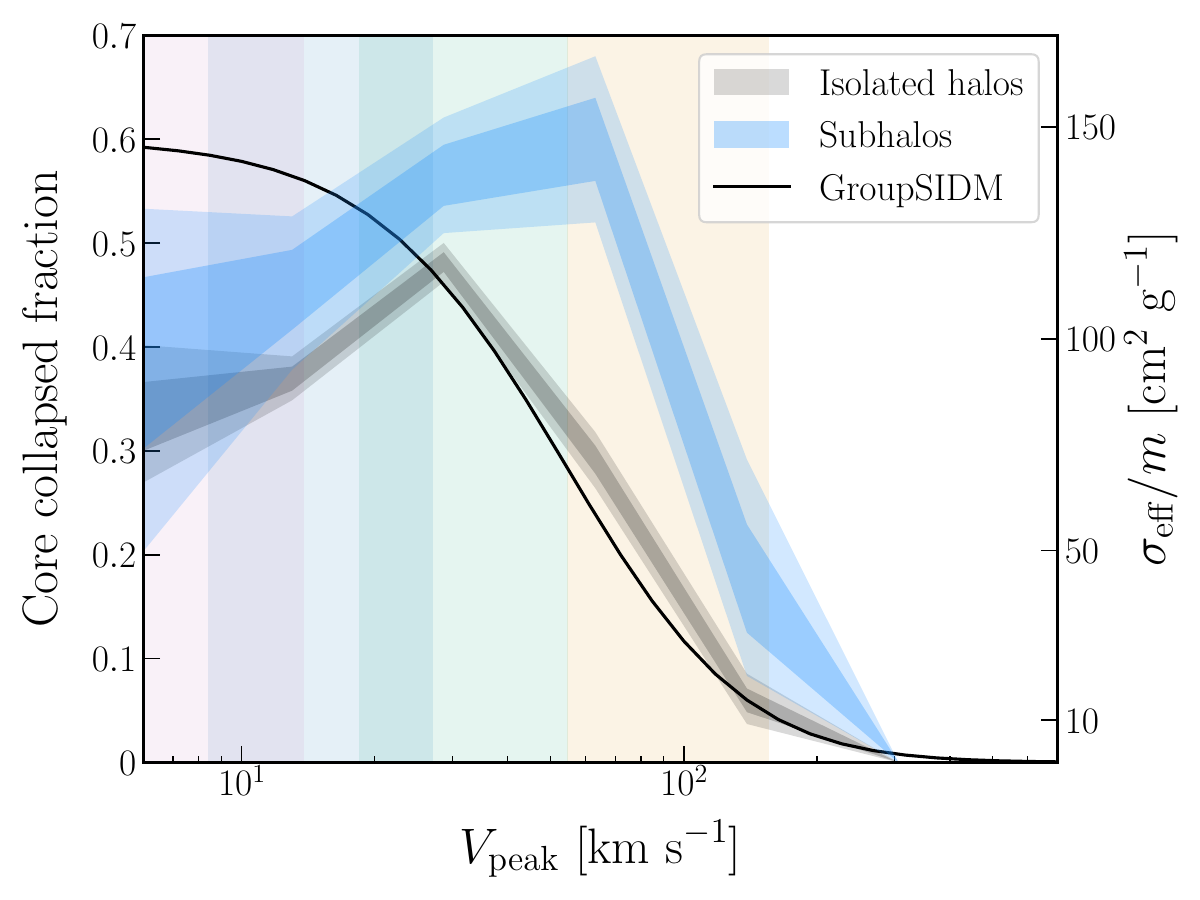}
   \caption{Core-collapsed fraction, defined as the fraction of halos with $\tau_0>0.75$ according to our parametric model predictions, for isolated halos (gray) and subhalos (blue) with $M_{\mathrm{vir}}>2000m_{\mathrm{part}}$. Results are combined for all GroupSIDM simulations as a function of peak maximum circular velocity, $V_{\mathrm{peak}}$. Dark (light) shaded bands show $68\%$ ($95\%$) confidence intervals from bootstrap resampling, which indicate that the decrease in the core-collapsed fractions at low $V_{\mathrm{peak}}$ is statistically significant. The GroupSIDM cross section as a function of $V_{\mathrm{max}}$ is overlaid in black with an arbitrary normalization. Vertical shaded bands show typical maximum circular velocities for (sub)halos in each suite, as in Figure~\ref{fig:xsec}. Note that $V_{\mathrm{peak}}\approx V_{\mathrm{max}}$ for isolated halos and that $V_{\mathrm{peak}}$ is typically significantly larger than $V_{\mathrm{max}}$ for subhalos.}
    \label{fig:f_cc}
\end{figure}

Next, Figure~\ref{fig:f_cc} shows the core-collapsed fraction, that is, the fraction of (sub)halos with $\tau_0>0.75$. We plot this fraction as a function of peak maximum circular velocity, $V_{\mathrm{peak}}$, which more directly correlates with initial halo concentration than $V_{\mathrm{max}}$ or mass~\citep{Lehmann151005651}, because the gravothermal evolution timescale is sensitive to initial concentration (e.g., \citealt{Essig180901144}). Consistent with the $\tau_0$ distributions in Figure~\ref{fig:tau_0}, we find that the core-collapsed fraction peaks at $V_{\mathrm{peak}}\approx 30~\mathrm{km\ s}^{-1}$ ($60~\mathrm{km\ s}^{-1}$), for isolated halos (subhalos) and declines toward lower $V_{\mathrm{peak}}$. This is the first demonstration of the turnover using cosmological simulations.

Following \cite{Ando240316633}, the peak in the core-collapsed fraction can be understood based on the interplay between the underlying SIDM cross section (which we overlay in black on Figure~\ref{fig:f_cc}) and the mass--concentration relation. In particular, the core-collapse timescale scales as~\citep{Balberg0110561}
\begin{equation}
    t_c\propto \frac{1}{(\sigma_{\rm eff}/m)r_s\rho_s^{3/2}}  
      \propto (\sigma_0/m)^{-1}M^{(n-1)/3}c^{(n-7)/2}, \label{eq:cc_scaling}
\end{equation}
where $\sigma_{\rm eff}/m \propto (\sigma_0/m)v^{-n}$ is the effective SIDM cross section per DM mass with velocity dependence $n$, $r_s$ ($\rho_s$) is the NFW profile scale radius (amplitude), $M$ ($c$) denotes halo mass (concentration), and we have used $v\sim V_{\mathrm{max}}\propto \rho_s^{1/2}r_s$, $\rho_s \propto c^3$, and $r_s \propto M^{1/3}/c$ to derive the second proportionality~\citep{Essig180901144,Nadler230601830}. Note that $\rho_s$ and $r_s$ describe an initial CDM (NFW) halo, which justifies using these scaling relations.

Thus, for high-mass halos where self-scattering probes the $n=4$ cross-section regime, $t_c\propto M c^{-3/2}$. Since concentration decreases with increasing halo mass, $t_c$ increases with halo mass in this regime. For low-mass halos where self-scattering probes the $n=0$ cross-section regime, $t_c\propto M^{-1/3} c^{-7/2}$. From \cite{Correa150200391}, $c\propto M^{-0.036}$ for halos with $M \lesssim 10^9~M_{\mathrm{\odot}}$ ($V_{\mathrm{peak}}\lesssim 25~\mathrm{km\ s}^{-1}$; \citealt{Nadler180905542}), we have $t_c\propto M^{-0.2}$, that is, $t_c$ increases as $M$ decreases. The turnover in the core-collapsed fraction therefore reflects the velocity dependence of the SIDM cross section, consistent with our measurements in Figure~\ref{fig:f_cc}. Note that the peak for subhalos is shifted to larger velocities relative to isolated halos because subhalos' gravothermal evolution is sensitive to both initial concentration (which correlates with $V_{\mathrm{peak}}$) and tidal evolution (which correlates with $V_{\mathrm{max}}$). The results in Figure~\ref{fig:f_cc} are similar when measured in terms of $V_{\mathrm{max}}$, and are broadly consistent with the core-collapsed fraction in \cite{Zeng231009910}, who studied a more extreme SIDM model. We leave a dedicated study of the factors that set $\tau_0$ for subhalos to future work.

\begin{figure*}[t!]
\centering
\includegraphics[trim={0 0.5cm 0 0cm},width=0.485\textwidth]{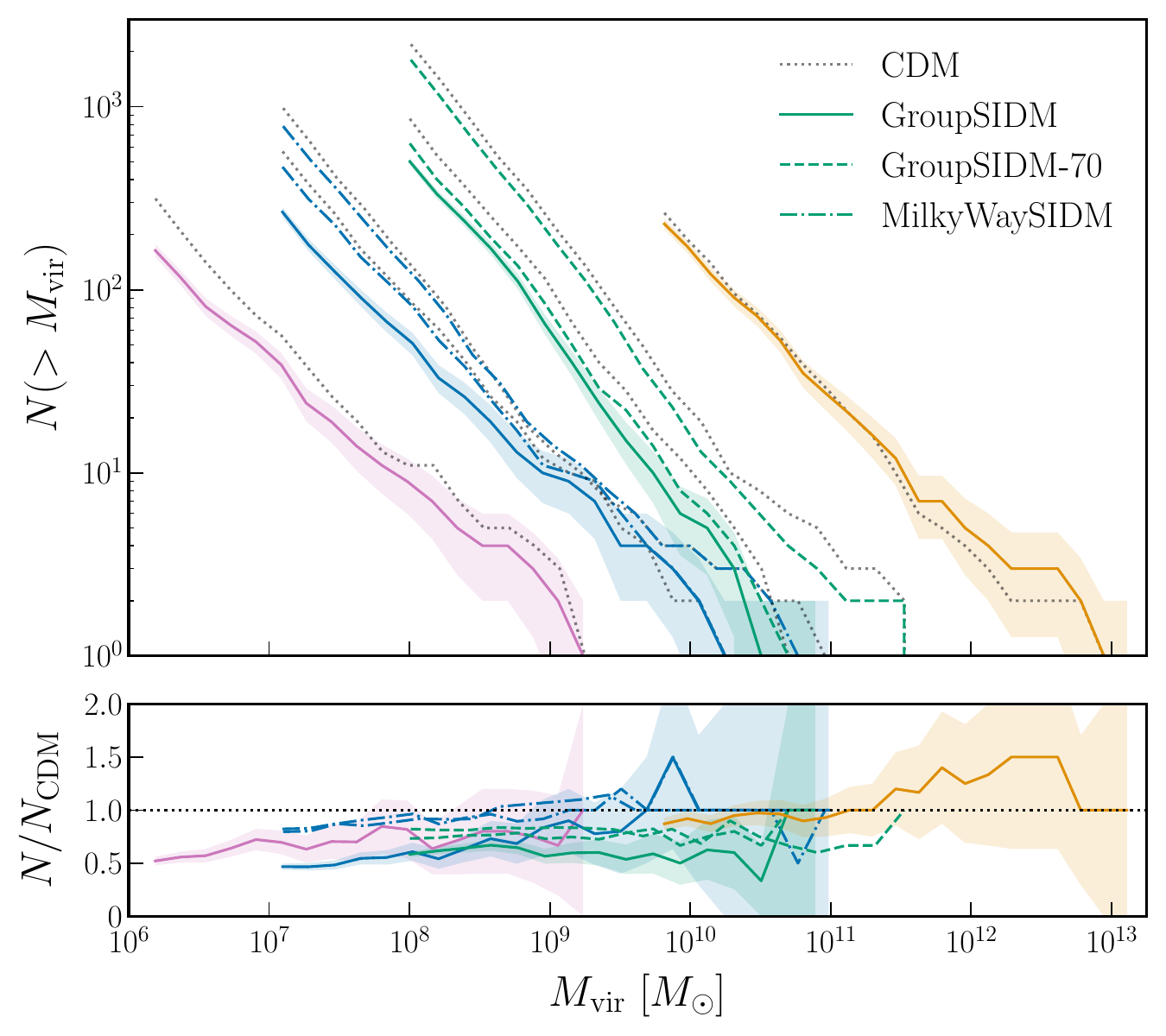}
\includegraphics[trim={0 0.5cm 0 0cm},width=0.485\textwidth]{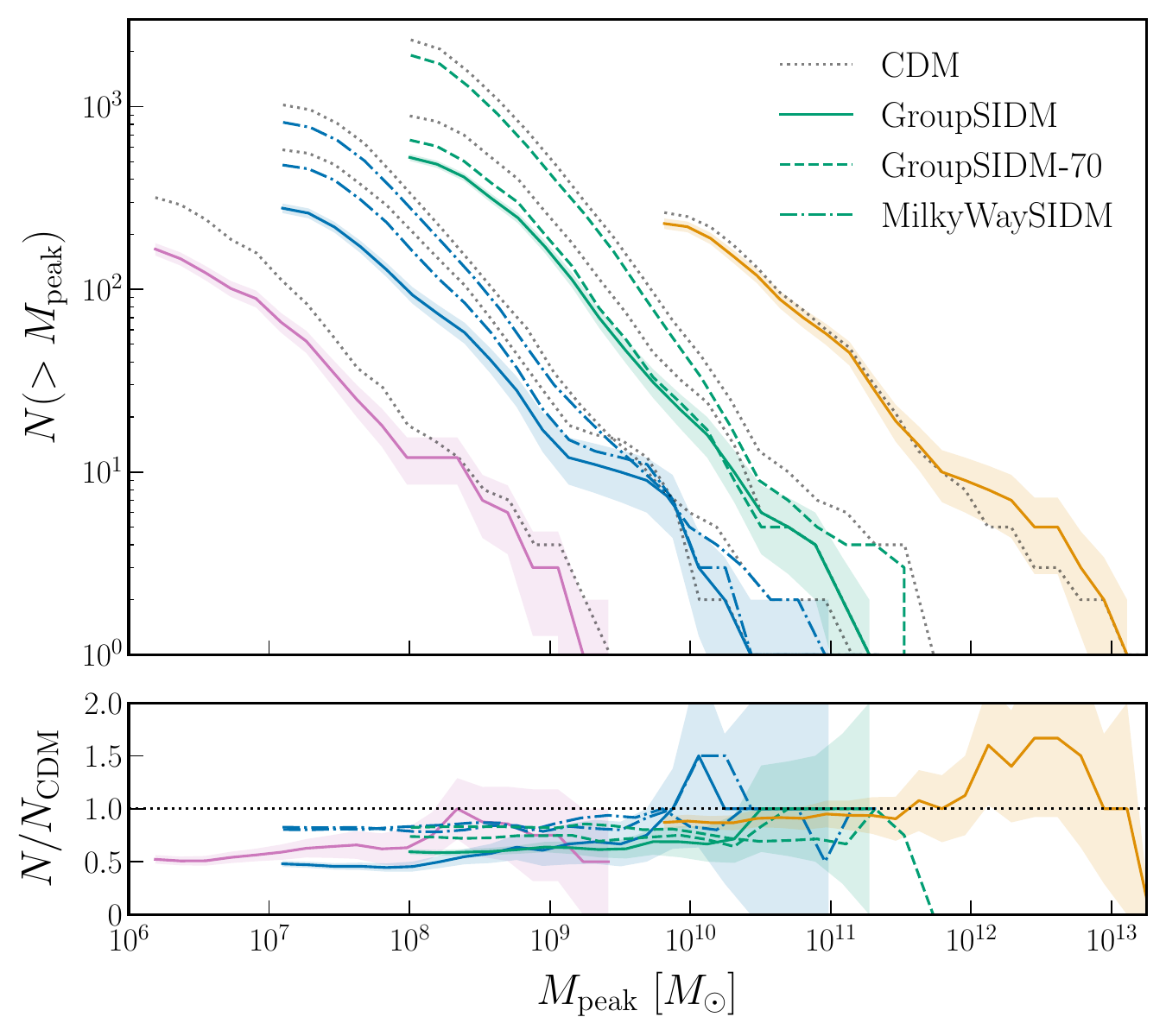}
    \caption{Cumulative SHMFs calculated using present-day virial mass (left panel) and peak virial mass (right panel) in SIDM (colored lines) and CDM (dotted black lines). SIDM results are shown for the GroupSIDM (solid), GroupSIDM-70 (dashed), and MilkyWaySIDM (dotted--dashed) models. Shaded bands show the $1\sigma$ Poisson uncertainty on the SIDM SHMFs, and bottom panels show ratios SIDM to CDM SHMFs. In both panels, we restrict to subhalos with $M_{\mathrm{vir}}>300 m_{\mathrm{part}}$.}
    \label{fig:shmf}
\end{figure*}

These results build on previous semianalytic predictions. In particular, our core-collapse timescale distributions are broadly consistent with the predictions from \cite{Shah230816342} and \cite{Ando240316633}, who, respectively applied semianalytic models for SIDM halo evolution to Symphony CDM simulations and to CDM subhalo populations generated by the SASHIMI structure formation model. The approach in \cite{Shah230816342} is similar to our parametric model application, but we emphasize that our model integrates over CDM (sub)halos' accretion histories rather than predicting their core-collapse timescales from properties at a single epoch. Our predictions thus capture the scatter in (sub)halo growth histories and differentiate between isolated halos and subhalos, which undergo gravothermal evolution at different rates due to tidal stripping (e.g., \citealt{Nishikawa190100499,Sameie190407872}).

\subsection{Subhalo Mass Functions}
\label{sec:shmf}

Figure~\ref{fig:shmf} shows cumulative subhalo mass functions (SHMFs) for each suite. SHMFs are evaluated using present-day (left panel) and peak (right panel) virial mass. For the LMC, MW, and Group suites, the GroupSIDM SHMF is suppressed by $\approx 50\%$ relative to CDM; this suppression is statistically significant given the Poisson uncertainty on our SHMF measurements. For the MW and Group hosts, this suppression is less severe for the MilkyWaySIDM and GroupSIDM-70 models, respectively; the former result is consistent with \cite{Yang221113768}.

The $\approx 50\%$ SHMF suppression in the GroupSIDM model is consistent with the MW zoom-in simulation results from \cite{Nadler200108754}, which were based on a lower-amplitude cross section and a different SIDM implementation but featured a similar cross-section amplitude at the subhalo infall velocity scale. Thus, the suppression is likely due to a combination of enhanced tidal disruption of cored subhalos and evaporation from subhalo--host halo interactions. We discuss prospects for constraining SIDM using this SHMF suppression signature in Section~\ref{sec:discussion}. Meanwhile, in the L-Cluster simulations, SIDM subhalo abundances are statistically consistent with CDM given the Poisson uncertainty on our SHMF measurements.\footnote{In the L-Cluster suite, the highest-mass subhalos are slightly \emph{more} abundant in SIDM than in CDM. While this is not statistically significant according to Figure~\ref{fig:shmf}, some previous SIDM simulations hint at a similar effect (e.g., \citealt{Fischer220502243}). A dedicated study is needed to understand the cause.}

We contextualize these SHMF results in several additional analyses. First, we reanalyze SHMFs using \textsc{Symfind} in Appendix~\ref{sec:symfind} to show that the main takeaways from the comparisons between CDM and SIDM in Figure~\ref{fig:shmf} are robust. Second, we present host halo density profiles in Appendix~\ref{sec:host_density}, since the SHMF differences (including the slight overabundance of high-mass SIDM subhalos in the L-Cluster suite) could partly be due to variations in tidal stripping rates caused by differences in the host potential between CDM and SIDM simulations of the same system. Finally, we show that isolated SIDM halo mass functions are nearly identical to CDM in Appendix~\ref{sec:isolated_hmf}, which implies that the differences in Figure~\ref{fig:shmf} are strictly due to postinfall evolution.

\begin{figure*}[t!]
\centering
\includegraphics[trim={0 0.5cm 0 0cm},width=\textwidth]{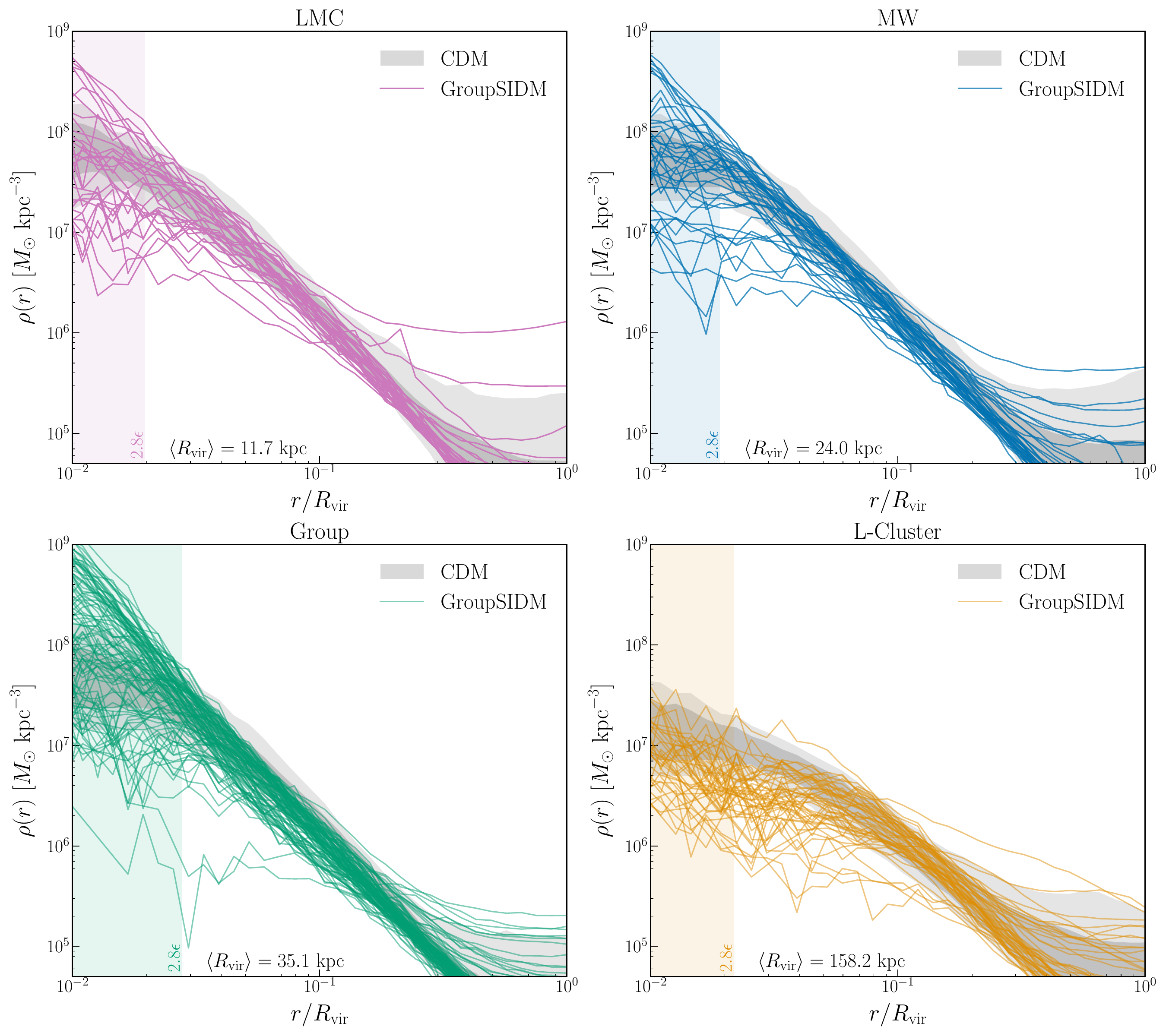}
    \caption{Density profiles for subhalos of in our LMC (Halo104; top left), MW (Halo004; top right), Group (Halo352; bottom left), and L-Cluster (Halo000; bottom right) zoom-ins. GroupSIDM results are shown by colored lines; dark (light) gray bands show $16\%$ ($84\%$) percentiles of the CDM density profile distribution in each host. In all panels, we restrict to subhalos with $M_{\mathrm{vir}}>2000 m_{\mathrm{part}}$. Vertical bands shade the region within $2.8\epsilon$ for each suite, determined using the average virial radius of subhalos above our resolution cut as indicated in each panel.}
    \label{fig:density_comparison}
\end{figure*}

\section{Subhalo Density Profiles}
\label{sec:density}

\begin{figure*}[t!]
\includegraphics[trim={0 0.5cm 0 0cm},width=0.5\textwidth]{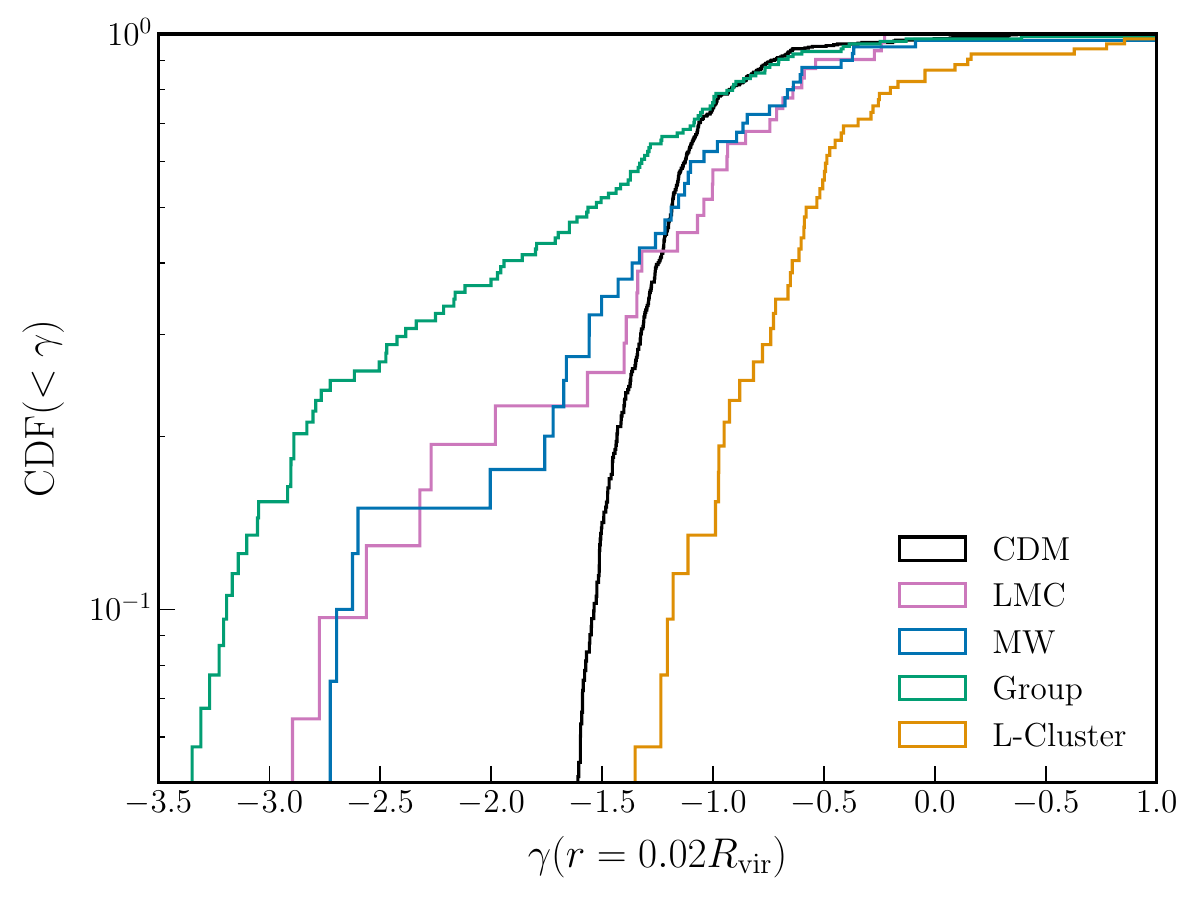}
\includegraphics[trim={0 0.5cm 0 0cm},width=0.5\textwidth]{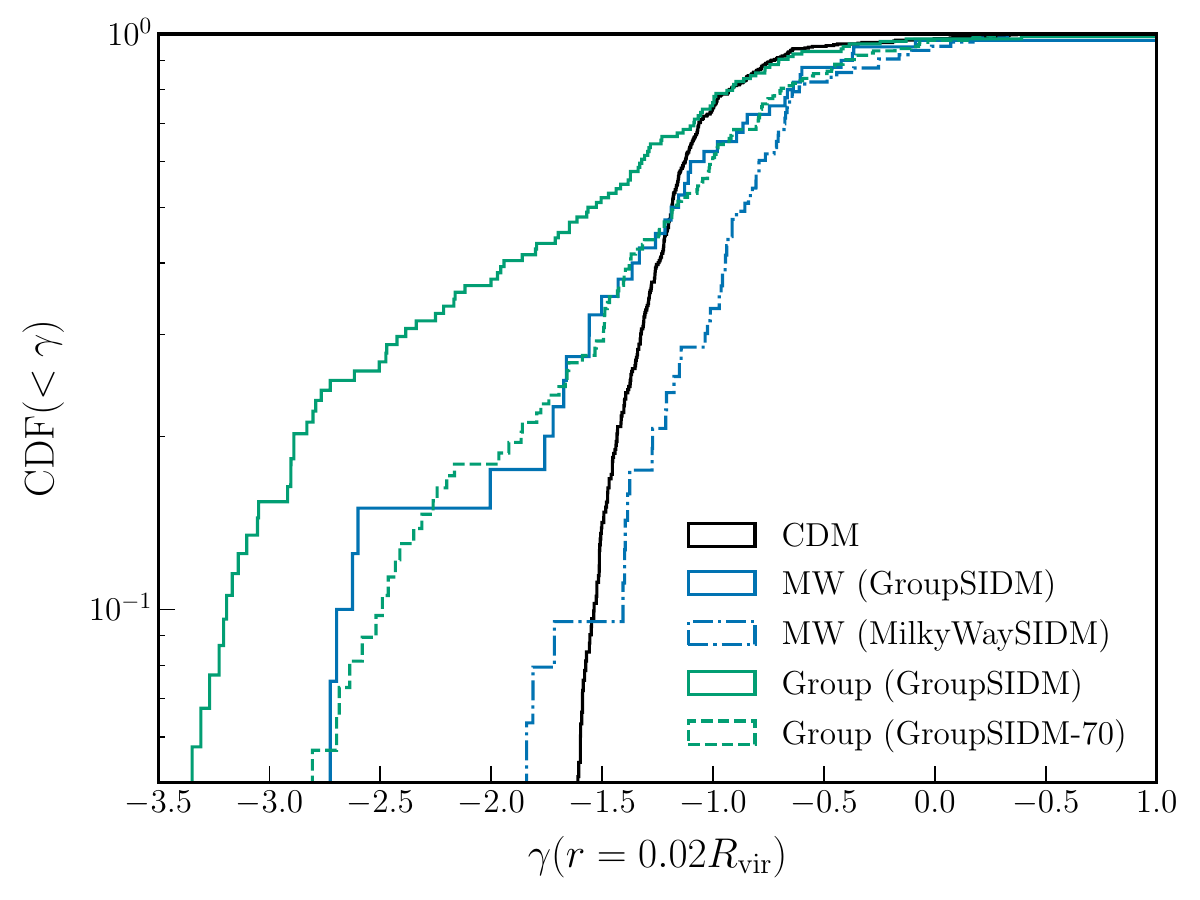}
    \caption{Logarithmic density profile slopes for the GroupSIDM (colored) and CDM (black) subhalos shown in Figure~\ref{fig:density_comparison} across all suites (left panel) and across different SIDM models in the MW and Group suites (right panel). Density profile slopes are evaluated at $0.02~R_{\mathrm{vir}}$, and CDM data from all hosts shown in Figure~\ref{fig:density_comparison} is combined. As in Figure~\ref{fig:density_comparison}, we restrict to subhalos with $M_{\mathrm{vir}}>2000 m_{\mathrm{part}}$. In the right panel, we compare GroupSIDM (solid), GroupSIDM-70 (dashed), and MilkyWaySIDM (dotted--dashed) results.}
    \label{fig:density_slopes}
\end{figure*}

We now study the density profiles of SIDM Concerto subhalos; we leave a study of isolated halo density profiles to future work. Figure~\ref{fig:density_comparison} compares density profiles for GroupSIDM subhalos (colored lines) and CDM subhalos (shaded bands) in each suite. We measure distances in units of each subhalo's virial radius because CDM profiles are approximately self-similar in these units \cite{Nadler220902675}. This normalization also highlights the relative scale within each subhalo by which SIDM core formation (or collapse) significantly affects halo structure. For cored halos, the central $\rho \sim r^0$ region can extend to $r \approx 0.1~R_{\mathrm{vir}}$. For core-collapsed halos, the central $\rho\sim r^{-2}$ region exceeds typical CDM density profiles near our convergence radius, which corresponds to $r\approx 0.02R_{\mathrm{vir}}$ across all suites. Note that core-collapsed subhalos often contain a large number of particles within the convergence radius due to their large enclosed inner masses.

Figure~\ref{fig:density_comparison} shows that GroupSIDM subhalo profiles are diversified relative to CDM at all host-mass scales. Both core-forming and core-collapsed subhalos are clearly visible in the LMC, MW, and Group panels, while the majority of surviving GroupSIDM subhalos of the L-Cluster host have lower central densities than their CDM counterparts.\footnote{We note, however, that some L-Cluster SIDM subhalos with rising inner density profiles could be in the mild core-collapse phase.} These findings are consistent with the corresponding $R_{\mathrm{max}}$--$V_{\mathrm{max}}$ relation results shown in Figure~\ref{fig:rmax_vmax}. 

Next, we fit cubic splines to subhalo profiles and measure logarithmic slopes $\gamma$ at $r=0.02R_{\mathrm{vir}}$, which is the smallest radius at which we robustly resolve density profiles in all suites. We combine all CDM results because the CDM distributions are statistically consistent across all host masses. The left panel of Figure~\ref{fig:density_slopes} shows that, for the GroupSIDM model, most subhalos of the LMC, MW, and Group hosts have steeper inner profiles than predicted in CDM, while L-Cluster subhalos generally have shallower inner profiles. At all host masses, the GroupSIDM inner slope distributions significantly differ from the CDM aggregate according to two-sample Kolmogorov-–Smirnov (KS) tests ($p=0.07$ and $0.004$ at the LMC and MW scales, respectively, and $p<10^{-5}$ at the Group and L-Cluster scales).

The right panel of Figure~\ref{fig:density_slopes} compares inner density profile slopes for different SIDM cross sections simulated in each host. Decreasing the cross-section amplitude (i.e., changing from GroupSIDM to GroupSIDM-70) or the velocity scale at which it transitions to a $v^0$ scaling (i.e., changing from GroupSIDM to MilkyWaySIDM) results in fewer core-collapsed subhalos and thus shifts the distribution toward less negative values. These shifts are statistically significant according to two-sample KS tests ($p=0.004$ in both cases). Thus, subhalos' inner density profile slopes are sensitive to both the amplitude and velocity transition scale of the underlying SIDM cross section.

\section{Discussion}
\label{sec:discussion}

We now discuss our results, focusing on prospects for SIDM constraints across the entire host halo-mass range we simulate (Section~\ref{sec:constraints}) and areas for future simulation work that build on SIDM Concerto (Section~\ref{sec:future}).

\subsection{Prospects for SIDM Constraints}
\label{sec:constraints}

The SIDM effects we have identified will inform cross-section constraints across a wide range of scattering velocities. Here we discuss observational prospects for probing SIDM based on subhalo abundances and density profiles at each host halo mass scale we simulate. In general, subhalo abundances probe scattering at both the relative subhalo--host velocity scale and at the characteristic internal velocity scales of subhalos, while density profiles mainly probe the latter. Note that, in addition to evaporation effects, cored halos are more likely to be tidally disrupted (e.g., \citealt{Errani221001131}).

\emph{LMC-mass hosts}. Surveys of satellite galaxies around the LMC itself (e.g., \citealt{Kallivayalil180501448,Drlica-Wagner210307476}) and around LMC-mass hosts throughout the Local Volume (e.g., \citealt{Carlin240917437}) will constrain their underlying SHMFs, and follow-up spectroscopic observations will probe these systems' inner DM densities. In particular, satellite luminosity function measurements will test for SHMF suppression due to SIDM, while central density measurements will probe the diversity of the underlying subhalo populations. The abundance and density profiles of LMC-associated subhalos can also potentially be probed through subhalo--stellar stream interactions, which are expected to be enhanced in the vicinity of the LMC system~\citep{Arora230915998}.

To leverage these observations, it will be important to develop models of SIDM satellite galaxy populations. These models could be built on SIDM Concerto by applying empirical and semianalytic models to our subhalo catalogs (e.g., following \citealt{Dooley170305321,Santos-Santos211101158}). In addition, hydrodynamic resimulations of LMC hosts will clarify the joint impact of baryons and SIDM on their subhalo populations (e.g., building on \citealt{Wang150304818,Jahn190702979}). Developing models of stellar disruption will be important to separate SHMF suppression due to SIDM from DM physics that suppresses the linear matter power spectrum like warm DM~\citep{Nadler241003635}, and also from enhanced tidal stripping due to the central LMC-mass galaxies (e.g., \citealt{Jahn190702979,Nadler210912120}).

\emph{MW-mass hosts}. Surveys of satellite galaxies around the MW itself (e.g., \citealt{Drlica-Wagner191203302}) and around MW-mass hosts throughout the Local Volume and low-redshift universe~\citep{Carlsten220300014,Mao240414498} will probe their underlying subhalo populations. Similar to LMC-mass satellites, interpreting these data will require models for SIDM satellite galaxy populations that include stellar disruption in the tidal field of MW-mass galaxies and account for the joint impact of baryons on the host and subhalo populations, which can be significant at the MW scale due to the high stellar-to-halo mass ratios of these hosts (e.g., \citealt{Rose220614830,Wang240801487}).

In addition to satellite galaxies, the MW's stellar stream population is a promising probe of DM physics including SIDM (for a review, see \citealt{Bonaca240519410}). Perturbations in nearby streams can be used to infer DM subhalo properties; for example, if the GD-1 gap-and-spur structure is due to an impact with a low-mass subhalo, the perturber must be extremely compact, with a density characteristic of core-collapsed SIDM subhalos~\citep{Zhang240919493}. Reconstructing stream perturber properties in this way can potentially constrain the distribution of inner density profiles for MW subhalos, which is sensitive to the SIDM cross section. To realize such constraints, it will be important to build toward analyses of the entire MW stellar stream population in CDM and SIDM models (e.g., \citealt{Lu250207781}).

\emph{Group-mass hosts}. Our Group hosts' halo masses are comparable to those of strongly lensed galaxies~\citep{Gavazzi0701589,Auger10072880}. Small-scale structure in such hosts and along the line of sight has been probed using flux ratio statistics~\citep{Hsueh190504182,Gilman190806983} and gravitational imaging~\citep{Vegetti09100760,Vegetti12013643,Hezaveh160101388}. Intriguingly, a perturber in the SDSSJ0946+1006 system detected via gravitational imaging is inferred to have an extremely high central density that is unusual in CDM~\citep{Minor201110627,Ballard230904535,Despali240712910}. A previous analysis of our Halo352 simulation in the GroupSIDM model shows that such dense perturbers are naturally produced in the strong, velocity-dependent SIDM models we simulate (\citealt{Nadler230601830}; also see \citealt{Zeng231009910}).

Upcoming surveys are expected to discover thousands of strongly lensed systems~\citep{Oguri10012037,Collett150702657}. Follow-up analyses will probe both the inner densities (typically on scales of $\sim 1~\mathrm{kpc}$; \citealt{Minor201110629,Despali240712910}) and abundances (down to $\sim 10^7~M_{\mathrm{\odot}}$; \citealt{Gilman190111031,Nierenberg230910101}) of low-mass (sub)halos. Our results demonstrate that both quantities are sensitive to the SIDM cross section. Robustly translating upcoming lensing measurements to SIDM constraints will require population analyses that account for selection effects~\citep{Sonnenfeld14101881}. SIDM Concerto represents an important step toward population-level SIDM strong lensing predictions.

\emph{Low-mass clusters}. Although we have simulated low-mass ($\sim 10^{14}~M_{\mathrm{\odot}}$) galaxy clusters, we will discuss prospects for SIDM constraints using these systems and larger ($\sim 10^{15}~M_{\mathrm{\odot}}$) clusters. Our cluster predictions can be tested by galaxy--galaxy strong lensing measurements. Recent work in this context has revealed a galaxy--galaxy strong lensing cross section that potentially exceeds CDM predictions (e.g., \citealt{Meneghetti200904471}). This signal is sensitive to cluster (sub)halo abundances and density profiles and can thus be used to test SIDM (e.g., \citealt{Yang210202375,Dutra240617024}).

SIDM Concerto and similar suites are also useful for deriving SIDM constraints from cluster density profiles, shapes, and mergers (e.g., \citealt{Harvey150307675,Sagunski200612515,Andrade201206611}; also see \citealt{Robertson221013474}). Weak lensing profiles are sensitive to the SIDM cross section at large scattering velocities, $\mathcal{O}(1000~\mathrm{km\ s}^{-1})$, which provides important constraints on velocity-dependent SIDM models (e.g., \citealt{Banerjee190612026,Bhattacharyya210608292,Adhikari240105788}). Developing simulations that capture the joint impact of baryons and SIDM for all of these observables will be important to improve current SIDM constraints from galaxy clusters (e.g., \citealt{Ragagnin240401383,Sirks240500140}).

\subsection{Building on SIDM Concerto}
\label{sec:future}

There are several exciting areas for future simulation work related to SIDM Concerto. First, hydrodynamic resimulations of SIDM Concerto hosts are timely. In particular, baryonic effects including adiabatic contraction and supernova feedback can alter SIDM-only predictions, particularly for (sub)halos with large stellar-to-halo mass ratios (e.g., \citealt{Kaplinghat13116524,Sameie180109682,Robles190301469,Rose220614830}). In hydrodynamic simulations, the joint impact of baryonic physics and SIDM on halo profiles depends on the details of the feedback prescription (e.g., \citealt{Robles170607514,Straight250116602}). Thus, it will be important to identify the degenerate and distinct signatures of SIDM and baryonic physics by simulating SIDM models in a wide range of feedback scenarios (e.g., building on \citealt{Burger210807358}).

Another natural extension is to include central galaxy potentials at each mass scale (e.g., following \citealt{Wang240801487}, who applied this method to the CDM Symphony MW suite). The impact of tidal stripping due to the central potential will vary over our host-mass range since the mass of the central galaxy and its mass ratio relative to the host vary as a function of host halo mass~\citep{Wechsler180403097}. In SIDM, we expect the central potential to enhance the disruption of cored subhalos (e.g., \citealt{Robles190301469}), while core-collapsed subhalos are likely more resilient to tidal disruption.

In parallel, it will be important to continue scrutinizing SIDM scattering algorithms and simulation analysis tools. While our SIDM implementation has been validated using controlled simulations~\citep{Yang220503392}, energy conservation for subhalos in the deep core-collapse phase remains challenging due to artificial heating~\citep{Fischer240300739,Zhang240919493}. Since this effect delays core collapse, our SIDM predictions likely underestimate the core-collapsed population. We plan to resimulate some of the SIDM Concerto systems with different gravity solvers, scattering algorithms, and analysis tools to assess these systematic modeling uncertainties. In terms of analysis tools, we have estimated uncertainties in our subhalo population results by comparing our \textsc{RCT} results to \textsc{Symfind} in Appendix~\ref{sec:symfind}. Following the recommendations in \cite{Kong250709799}, future SIDM Concerto analyses can combine the strengths of \textsc{RCT} and \textsc{Symfind} to more accurately track subhalos in different stages of gravothermal evolution. Nonetheless, we emphasize that our main results related to internal halo structure (e.g., the turnover in the core-collapsed fraction at low $V_{\mathrm{peak}}$) hold for isolated halos as well as subhalos, which lends confidence to our conclusions.

Finally, we anticipate that combining controlled simulations with SIDM Concerto merger trees, building on the work of \cite{Zhang240919493}, is a compelling avenue for further work. By initializing subhalos according to their properties in cosmological simulations at infall and subsequently evolving them at extremely high resolution in an analytic gravitational potential, this method helps address some of the numerical uncertainties associated with $N$-body simulations while retaining the cosmological context of zoom-in simulations. Note that the Symphony and Milky Way-est hosts we draw on span a range of environments, such that lower-mass hosts are found in more underdense large-scale environments~\citep{Nadler220902675}; this environmental variation may also influence SIDM predictions. We look forward to combining these tools across the entire SIDM Concerto host halo mass range to further refine SIDM predictions.


\section{Conclusions}
\label{sec:conclusions}

We have presented SIDM Concerto: $14$ high-resolution cosmological DM-only zoom-ins run in both CDM and strong, velocity-dependent SIDM models favored by recent small-scale structure anomalies. This compilation builds on previous CDM~\citep{Nadler220902675,Buch240408043} and SIDM~\citep{Nadler230601830,Nadler241213065,Yang221113768} simulation suites. SIDM Concerto enables studies of SIDM (sub)halo populations over an unprecedented dynamic range spanning host halo masses from $\sim 10^{11}$ to $10^{14}~M_{\mathrm{\odot}}$ and subhalo masses from $\sim 10^6$ to $10^{13}~M_{\mathrm{\odot}}$. 

Our main findings, which apply to the SIDM models shown in Figure~\ref{fig:xsec}, are as follows:
\begin{enumerate}
    \item The fraction of core-collapsed (sub)halos peaks at a velocity scale set by the SIDM cross section (Figure~\ref{fig:f_cc}). This is the first demonstration of the core-collapsed fraction turnover using cosmological simulations.
    \item SIDM subhalo abundances are suppressed by $\approx 50\%$ relative to CDM in LMC, MW, and Group hosts but are not altered in low-mass clusters (Figure~\ref{fig:shmf}).
    \item (Sub)halos' inner density profile slopes are sensitive to the SIDM cross-section amplitude and the velocity at which it transitions from a $v^{-4}$ to $v^0$ scaling (Figure~\ref{fig:density_slopes}).
\end{enumerate}

This work serves as a compilation and first data release of SIDM Concerto. Given the upcoming influx of small-scale structure data (e.g., \citealt{Bechtol220307354,Chakrabarti220306200}), we anticipate that these simulations will provide a foundation to search for evidence of DM self-interactions using observations of isolated and satellite galaxies, strong lens systems, stellar streams, and beyond.


\section*{Acknowledgements}

(Sub)halo catalogs, particle snapshots, and parametric model data are distributed on Zenodo at \url{https://doi.org/10.5281/zenodo.14933624}. Our modified \textsc{Symfind} code is available at \url{https://github.com/DemaoK/Concerto}.

This work was supported by the US Department of Energy under grant No.\ de-sc0008541 (H.-B.Y.). Simulations were conducted through Carnegie's partnership in the Resnick High Performance Computing Center, a facility supported by Resnick Sustainability Institute at Caltech. Analyses were performed using the clusters and data storage resources of the HPCC at UCR, which were funded
by the NSF (MRI-2215705, MRI-1429826) and
NIH (1S10OD016290-01A1). This
work used data from the Symphony and Milky Way-est suites, hosted at \url{https://web.stanford.edu/group/gfc/gfcsims/}, which were supported by the Kavli Institute for Particle Astrophysics and Cosmology at Stanford University, SLAC National Accelerator Laboratory, and the US DOE under contract No.\ DE-AC02-76SF00515 to SLAC.

\software{
{\sc consistent-trees} \citep{Behroozi11104370},
\textsc{Helpers} (\http{bitbucket.org/yymao/helpers/src/master/}),
\textsc{Jupyter} (\http{jupyter.org}),
\textsc{Matplotlib} \citep{matplotlib},
\textsc{meshoid} (\url{https://github.com/mikegrudic/meshoid}),
\textsc{NumPy} \citep{numpy},
\textsc{pynbody} \citep{pynbody},
{\sc Rockstar} \citep{Behroozi11104372},
\textsc{SciPy} \citep{scipy},
\textsc{Seaborn} (\https{seaborn.pydata.org}).
}

\bibliographystyle{yahapj2}
\bibliography{references,software}


\appendix

\section{Convergence Tests}
\label{sec:convergence}

To test for convergence, we rerun every simulation with one fewer \textsc{MUSIC} refinement region. For the LMC, MW, Group, and L-Cluster suites, these low-resolution (LR) simulations have $m_{\mathrm{part,LR}}=[5\times 10^4,~ 4\times 10^5,~ 3.2\times 10^6,~ 2.2\times 10^8]~M_{\mathrm{\odot}}$ and $\epsilon_{\mathrm{LR}}=[80,~ 170,~ 360,~ 1200]~\mathrm{pc}\ h^{-1}$. Here we show that SIDM $R_{\mathrm{max}}$--$V_{\mathrm{max}}$ relations, SHMFs, and density profiles are converged down to either the $2000m_{\mathrm{part,LR}}$ or $300m_{\mathrm{part,LR}}$ limit, lending confidence to our main results. Note that CDM convergence properties for a subset of our simulations were studied in \cite{Nadler220902675}, \cite{Yang221113768}, \cite{Buch240408043}, and \cite{Nadler241213065}.

\subsection{$R_{\mathrm{max}}$--$V_{\mathrm{max}}$ Relations}
\label{sec:rmax_vmax_convergence}

\begin{figure*}[t!]
\centering
\includegraphics[width=\textwidth]{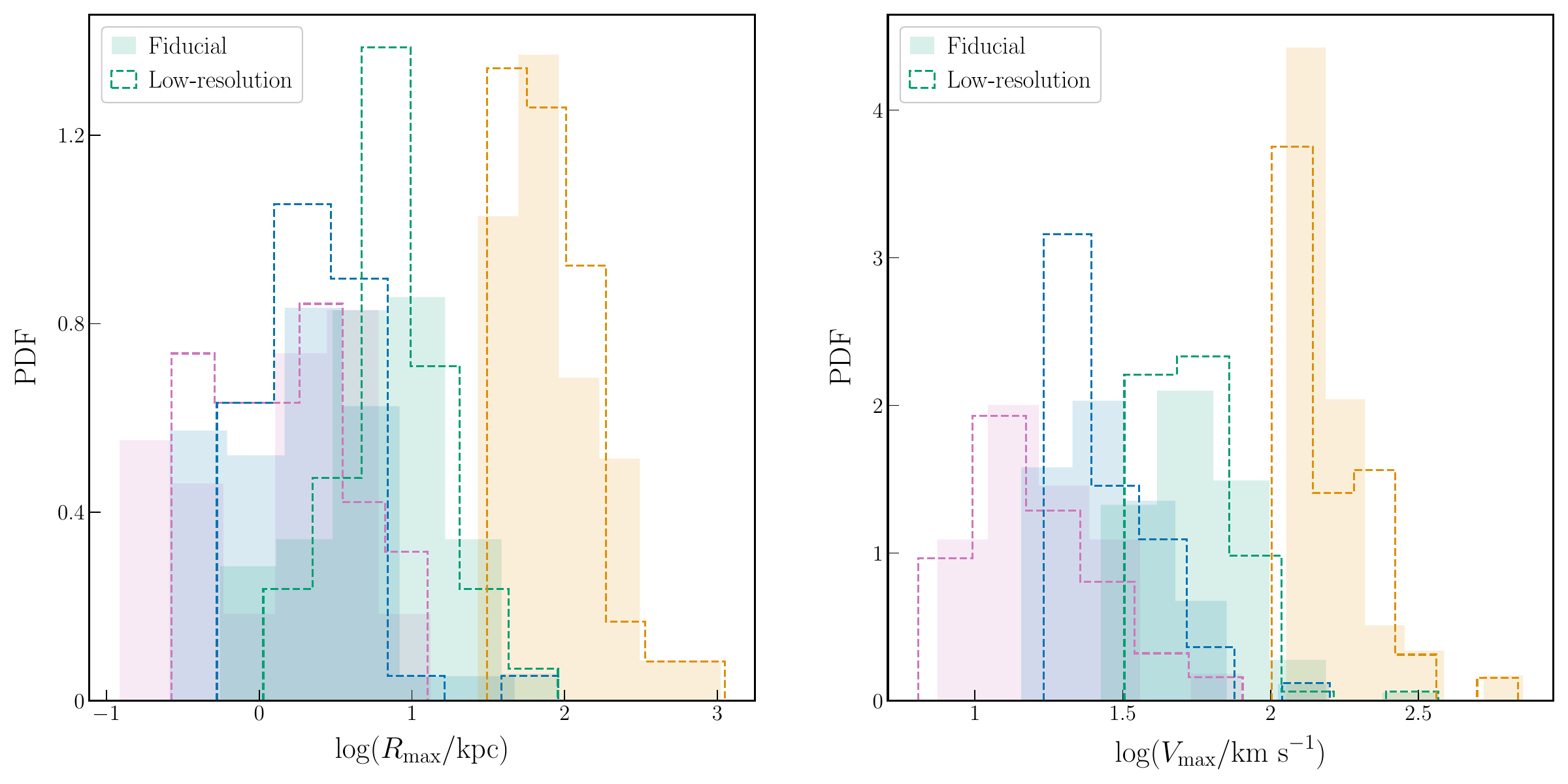}
    \caption{Normalized distributions of $R_{\mathrm{max}}$ (left panel) and $V_{\mathrm{max}}$ (right panel) for isolated halos with $M_{\mathrm{vir}}>2000m_{\mathrm{part,LR}}$ in each GroupSIDM simulation run at our standard-resolution (shaded) and from our LR resimulations (dashed).}
    \label{fig:rmax_vmax_convergence}
\end{figure*}

Figure~\ref{fig:rmax_vmax_convergence} compares the distribution of $R_{\mathrm{max}}$ and $V_{\mathrm{max}}$ for isolated halos in our GroupSIDM simulations with $M_{\mathrm{vir}}>2000m_{\mathrm{part,LR}}$ in our fiducial-resolution and LR simulations. We restrict to isolated halos because there are too few subhalos above the LR resolution threshold to robustly test for convergence. Across all suites, we find that the fiducial and LR distributions are statistically consistent according to two-sample KS tests. Thus, isolated halos' structural properties are well converged above a $2000$-particle limit.

\subsection{Subhalo Mass Functions}
\label{sec:hmf_convergence}

\begin{figure*}[t!]
\centering
\includegraphics[trim={0 0.5cm 0 0cm},width=0.485\textwidth]{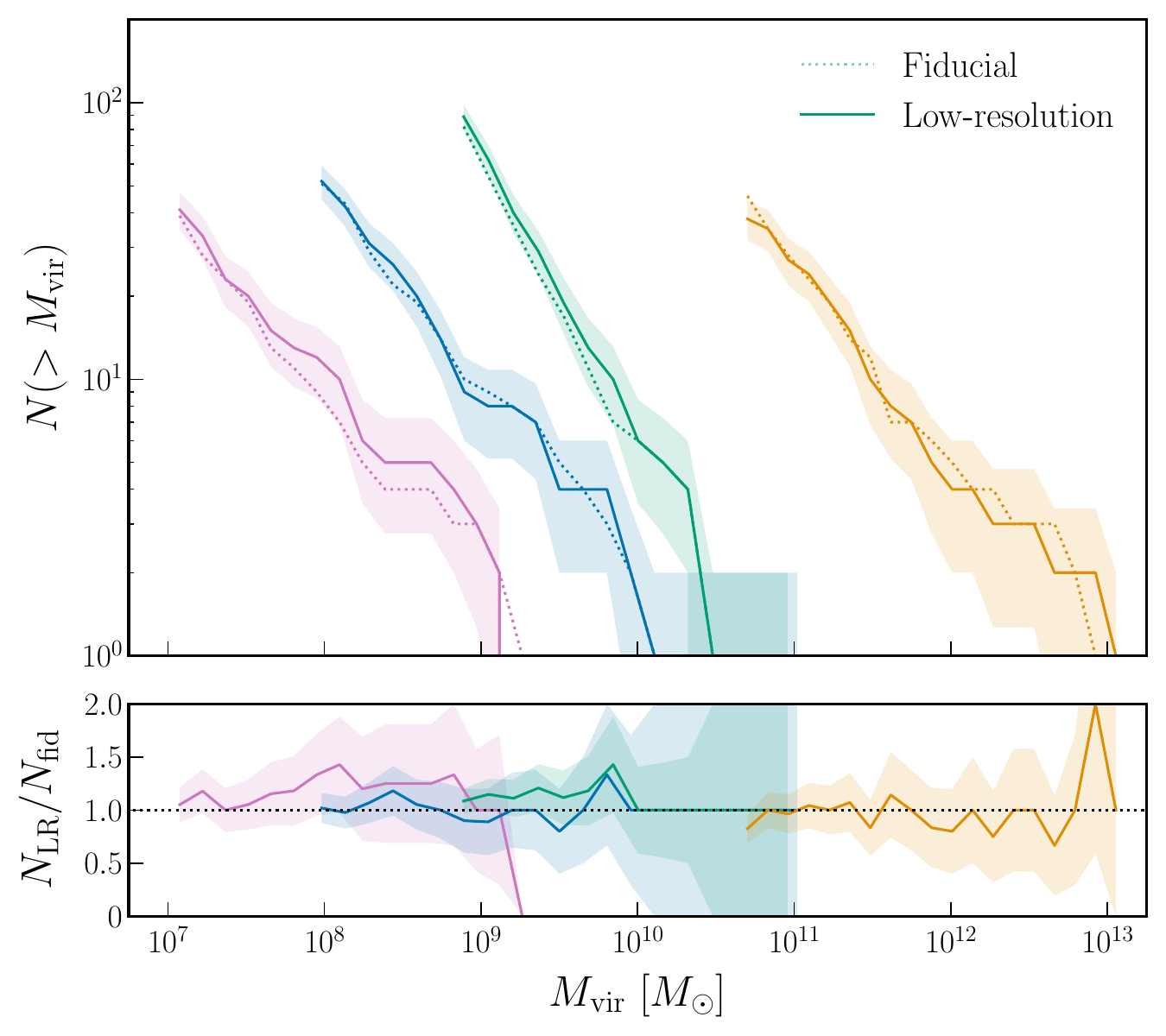}
\includegraphics[trim={0 0.5cm 0 0cm},width=0.485\textwidth]{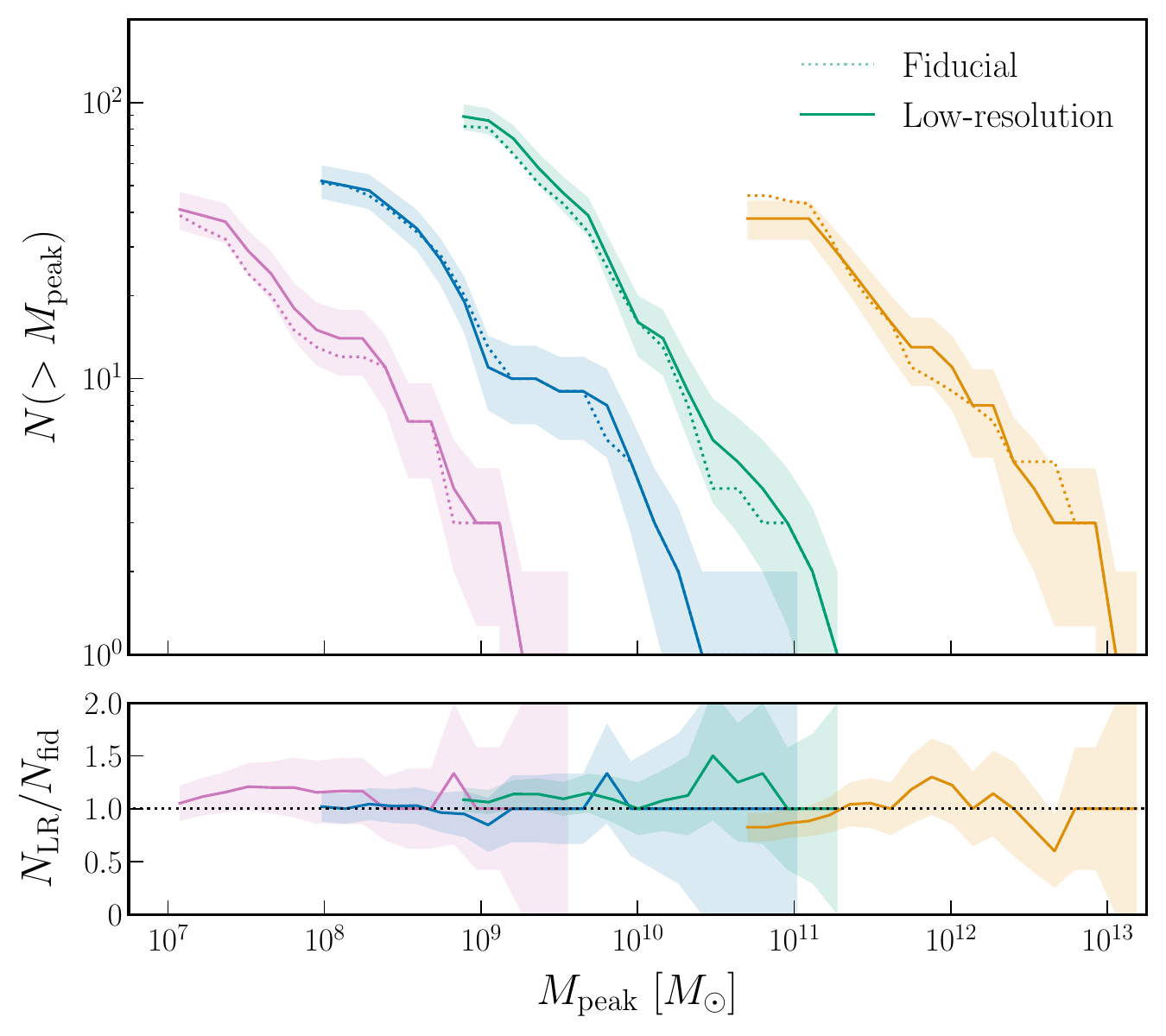}
    \caption{Cumulative $M_{\mathrm{vir}}$ (left panel) and $M_{\mathrm{peak}}$ (right panel) SHMFs for our standard-resolution (dotted) and LR (solid) GroupSIDM simulations. Bottom panels show the ratio of the LR to fiducial SHMFs. In both panels, we restrict to subhalos with $M_{\mathrm{vir}}>300 m_{\mathrm{part,LR}}$.}
    \label{fig:shmf_convergence}
\end{figure*}

Figure~\ref{fig:shmf_convergence} compares cumulative SHMFs from our standard-resolution and LR GroupSIDM simulations. For both the $M_{\mathrm{vir}}$ and $M_{\mathrm{peak}}$ functions, we find that SHMFs are statistically consistent above $300m_{\mathrm{part,LR}}$ across all host halo masses we simulate. Thus, subhalo abundances are well converged above a $300$-particle limit. We also find that isolated HMFs are well converged above this limit.

\subsection{Subhalo Density Profiles}
\label{sec:density_convergence}

\begin{figure*}[t!]
\centering
\includegraphics[trim={0 0.5cm 0 0cm},width=\textwidth]{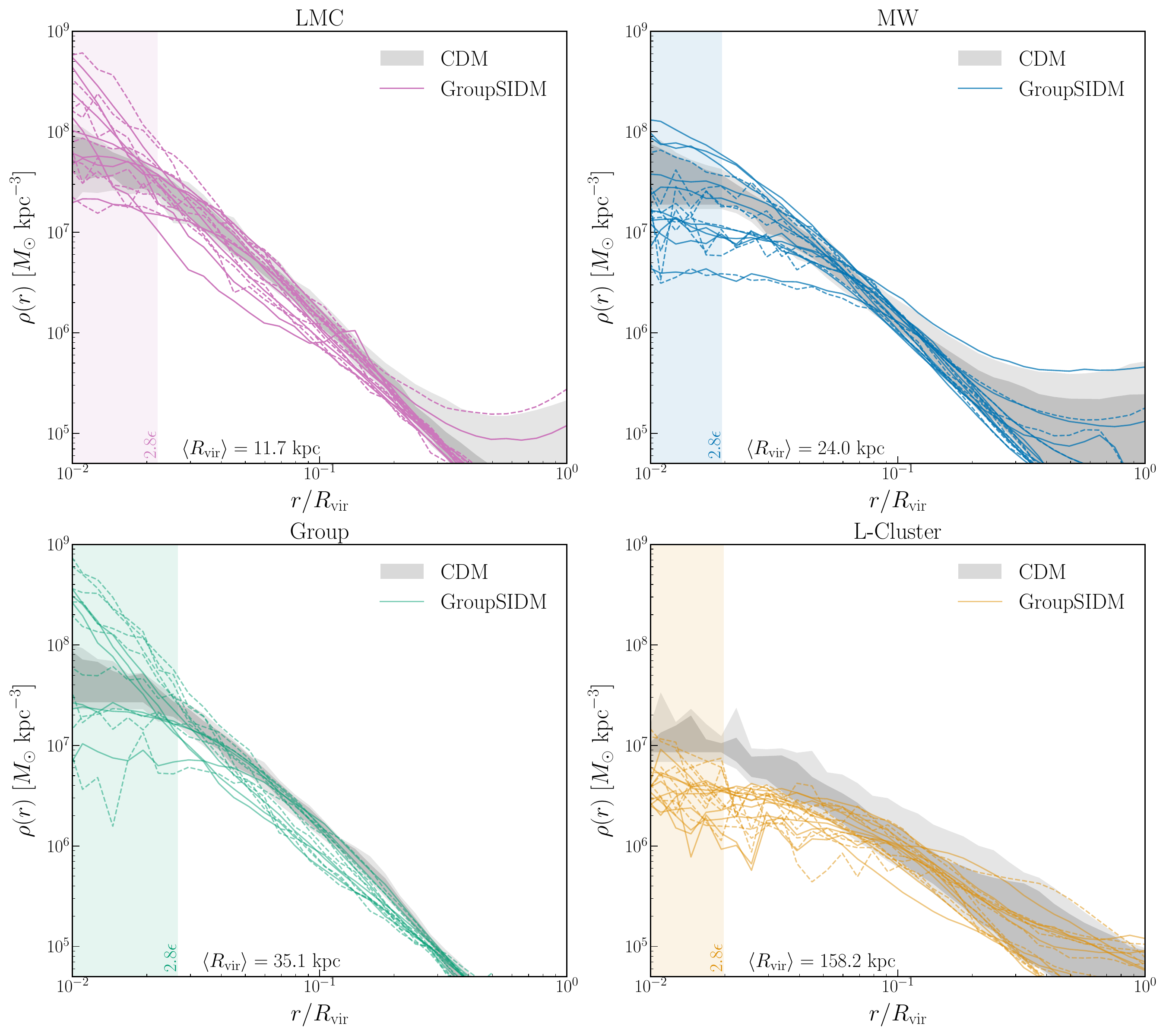}
    \caption{Same as Figure~\ref{fig:density_comparison}, but comparing subhalo density profiles in our fiducial-resolution (solid) and LR (dashed) GroupSIDM simulations. In all panels, we restrict to subhalos with $M_{\mathrm{vir}}>2000 m_{\mathrm{part,LR}}$. Vertical bands shade the region within $2.8\epsilon_{\mathrm{LR}}$ for each suite, determined using the average virial radius of subhalos above our resolution cut indicated in each panel.}
    \label{fig:density_comparison_LR}
\end{figure*}

Figure~\ref{fig:density_comparison_LR} compares GroupSIDM density profiles for subhalos with $M_{\mathrm{vir}}>2000m_{\mathrm{part,LR}}$ in our fiducial-resolution and LR simulations. At each host-mass scale, the density profile distribution is consistent between resolution levels. We test this quantitatively using two-sample KS tests to compare the density profile slopes at $0.02R_{\mathrm{vir}}$ at each host mass; none of these tests yield significant differences between the resolution levels, confirming that our fiducial results are converged down to a $2000$-particle limit. 

\section{Comparison between Rockstar and Symfind SHMFs}
\label{sec:symfind}

\begin{figure*}[t!]
\centering
\includegraphics[width=0.485\textwidth]{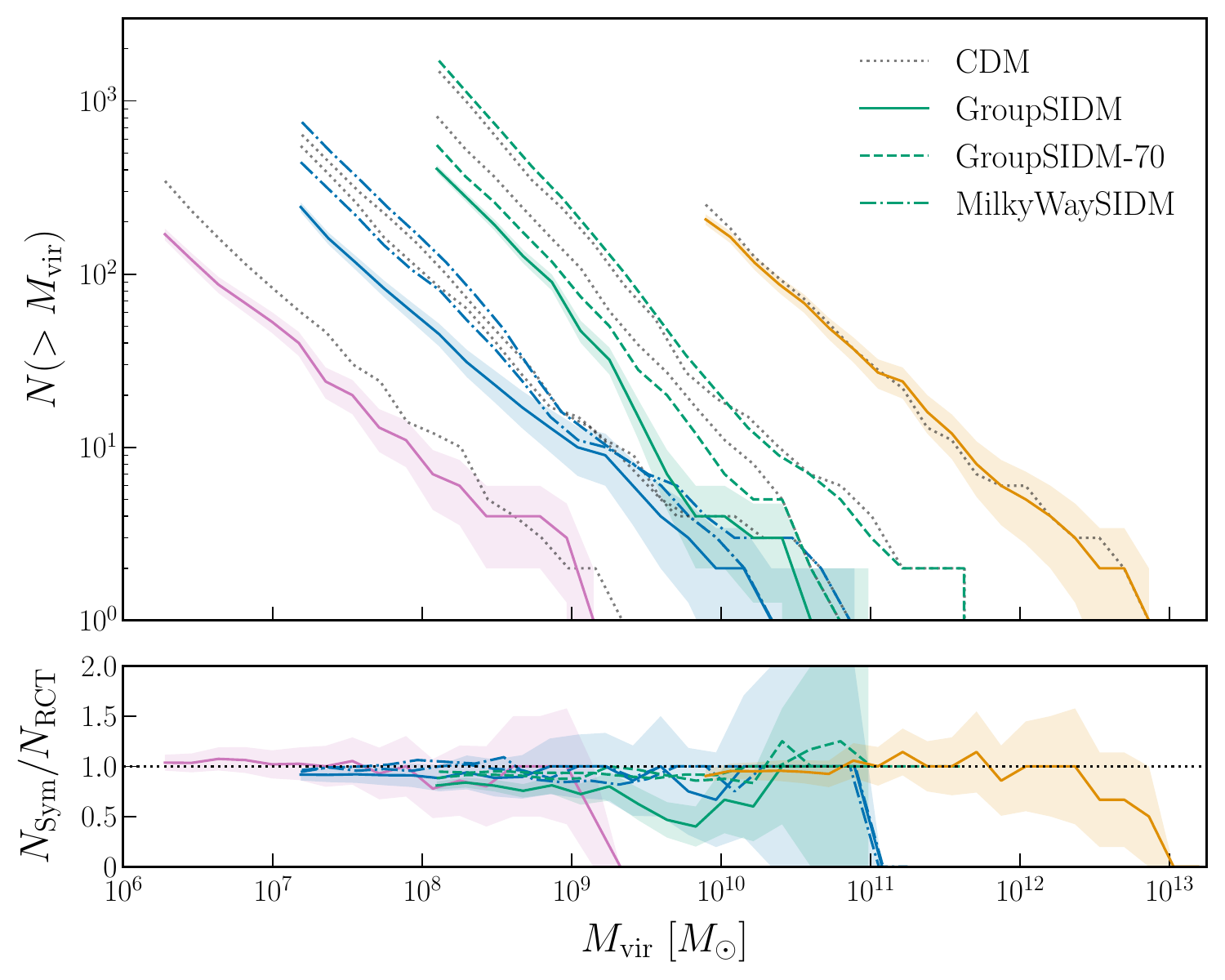}
\includegraphics[width=0.485\textwidth]{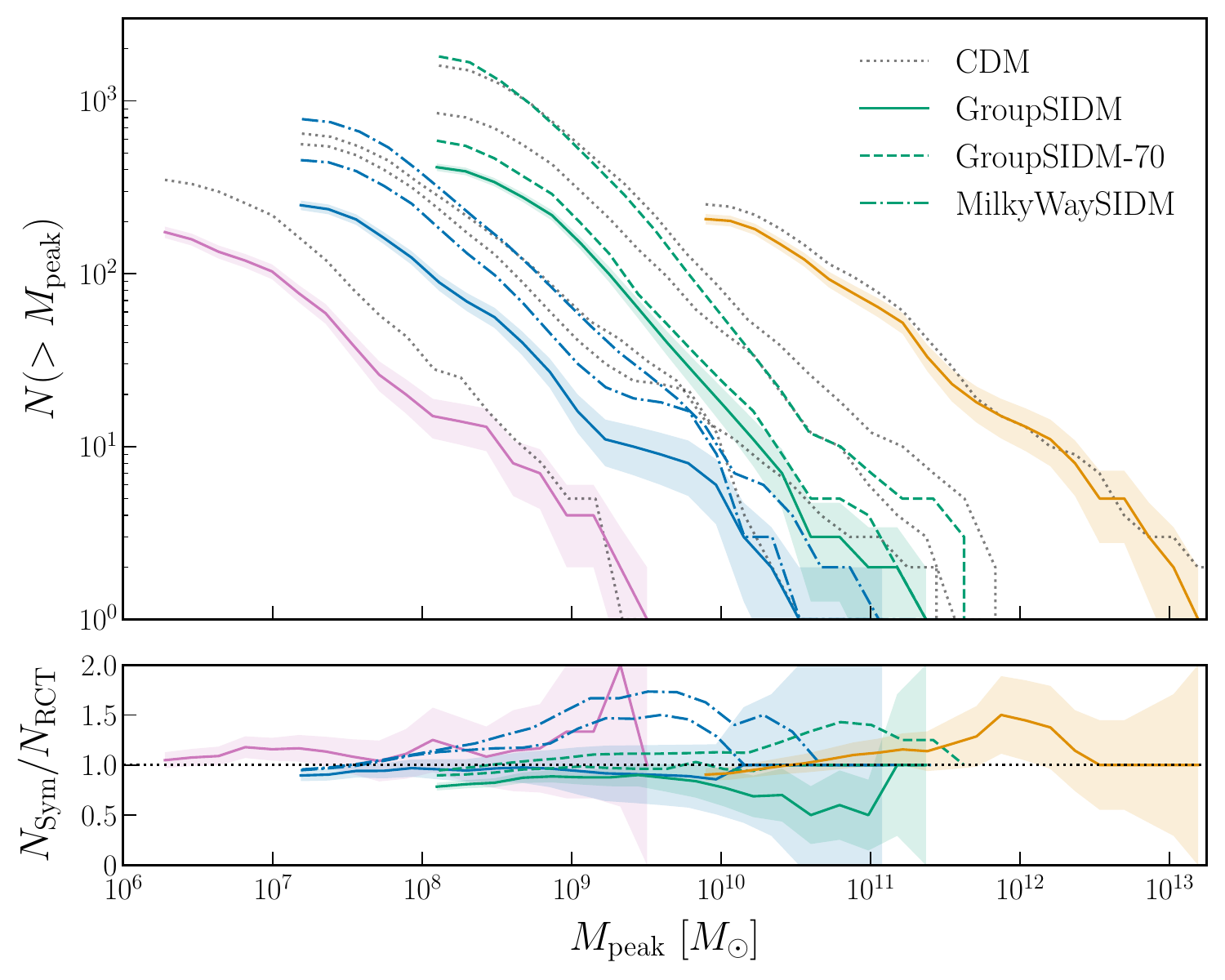}
    \caption{Cumulative subhalo mass functions calculated using \textsc{Symfind}. As in Figure~\ref{fig:shmf}, we plot present-day virial mass (left panel) and peak virial mass (right panel) in SIDM (colored lines) and CDM (dotted black lines). SIDM results are shown for the GroupSIDM (solid), GroupSIDM-70 (dashed), and MilkyWaySIDM (dotted--dashed) models. Shaded bands show the $1\sigma$ Poisson uncertainty on the SIDM subhalo mass functions. Bottom panels show the ratio of \textsc{Symfind} to \textsc{RCT} SIDM subhalo mass functions. In both panels, we restrict to subhalos with $M_{\mathrm{vir}}>300 m_{\mathrm{part}}$.}
    \label{fig:shmf_symfind}
\end{figure*}

Figure~\ref{fig:shmf_symfind} shows SHMFs calculated using \textsc{Symfind} (top panels) and the ratio of the \textsc{Symfind} to our fiducial \textsc{RCT} results. We refer the reader to \cite{Kong250709799} for a detailed discussion of applying particle-tracking-based subhalo finders to SIDM simulations. Here we focus on the main takeaways relevant for the interpretation of SIDM Concerto SHMF results.

First, for the $M_{\mathrm{vir}}$ SHMFs, all \textsc{Symfind} and \textsc{RCT} results are statistically consistent at the lowest subhalo masses we resolve in each suite, with the exception of Halo352 GroupSIDM, which we discuss below. There is a general trend for $N_{\mathrm{Sym}}/N_{\mathrm{RCT}}$ to decrease toward higher subhalo masses. \cite{Kong250709799} show that---for the velocity-dependent SIDM models with large cross section amplitudes that we study---\textsc{Symfind} tracks SIDM subhalos more accurately than \textsc{RCT} in the MilkyWaySIDM model. Conversely, when the turnover velocity is higher, as in the GroupSIDM model, \textsc{RCT} outperforms \textsc{Symfind}. In the GroupSIDM model, subhalo ``core'' particles tagged by \textsc{Symfind} tend to be kicked out to less bound orbits due to self-interactions; these particles can then be tidally stripped, causing \textsc{Symfind} to underperform. This preferentially affects cored subhalos, although the results depend on the specific tidal evolution history of each system. Meanwhile, core particles remain more bound in both the MilkyWaySIDM and GroupSIDM-70 models compared to the GroupSIDM model. In the former case, the lower turnover velocity reduces the strength of the cross section for massive cored subhalos, and core particle diffusion is suppressed. In the latter case, the lower cross-section amplitude yields subhalos with smaller density cores, such that core particles are more tightly bound.

Next, for the $M_{\mathrm{peak}}$ SHMFs, \textsc{Symfind} tends to resolve significantly more subhalos than \textsc{RCT} across all simulations except Halo352 GroupSIDM. This trend is not highly sensitive to $M_{\mathrm{peak}}$ within each simulation, consistent with the CDM \textsc{Symfind} results from \cite{Mansfield230810926}. The difference is due to \textsc{Symfind} more robustly tracking subhalos at a given $M_{\mathrm{peak}}$ as they are tidally stripped. Combined with the $M_{\mathrm{vir}}$ SHMF result, we conclude that our \textsc{RCT} subhalo catalogs are largely complete down to a fixed $M_{\mathrm{vir}}$ threshold and that they miss at most $\approx 50\%$ of highly stripped subhalos down to a fixed $M_{\mathrm{peak}}$ threshold. These missing objects likely do not have prominent cores, since \textsc{Symfind} often loses track of such subhalos as core particles escape during tidal stripping. Thus, our main results are conservative in the sense that they underestimate the SIDM core-collapsed fraction. Furthermore, all of our main analyses that rely on \textsc{RCT} impose $M_{\mathrm{vir}}$ cuts, above which subhalo abundances are robustly measured.

\section{Host Halo Density Profiles}
\label{sec:host_density}

\begin{figure*}[t!]
\includegraphics[trim={0 0.5cm 0 0cm},width=0.5\textwidth]{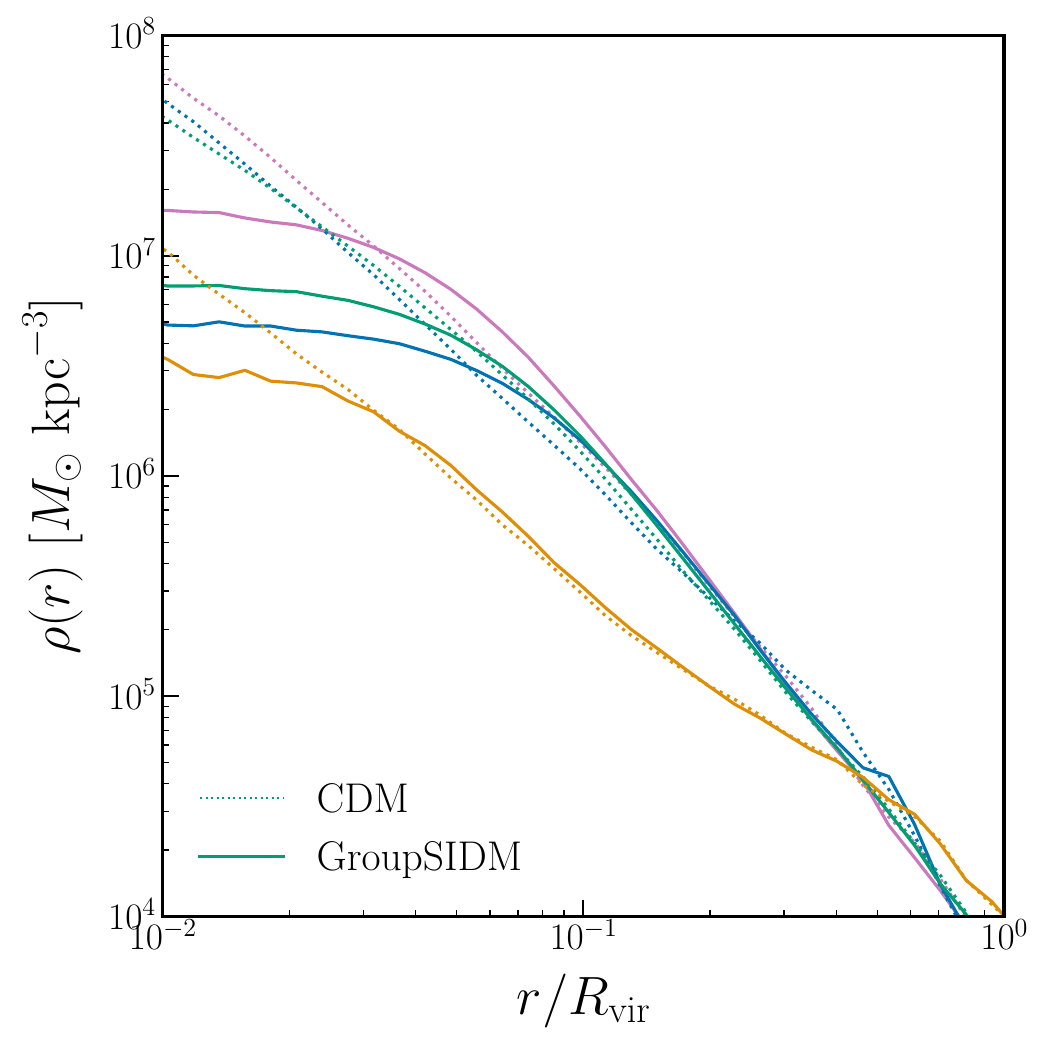}
\includegraphics[trim={0 0.5cm 0 0cm},width=0.5\textwidth]{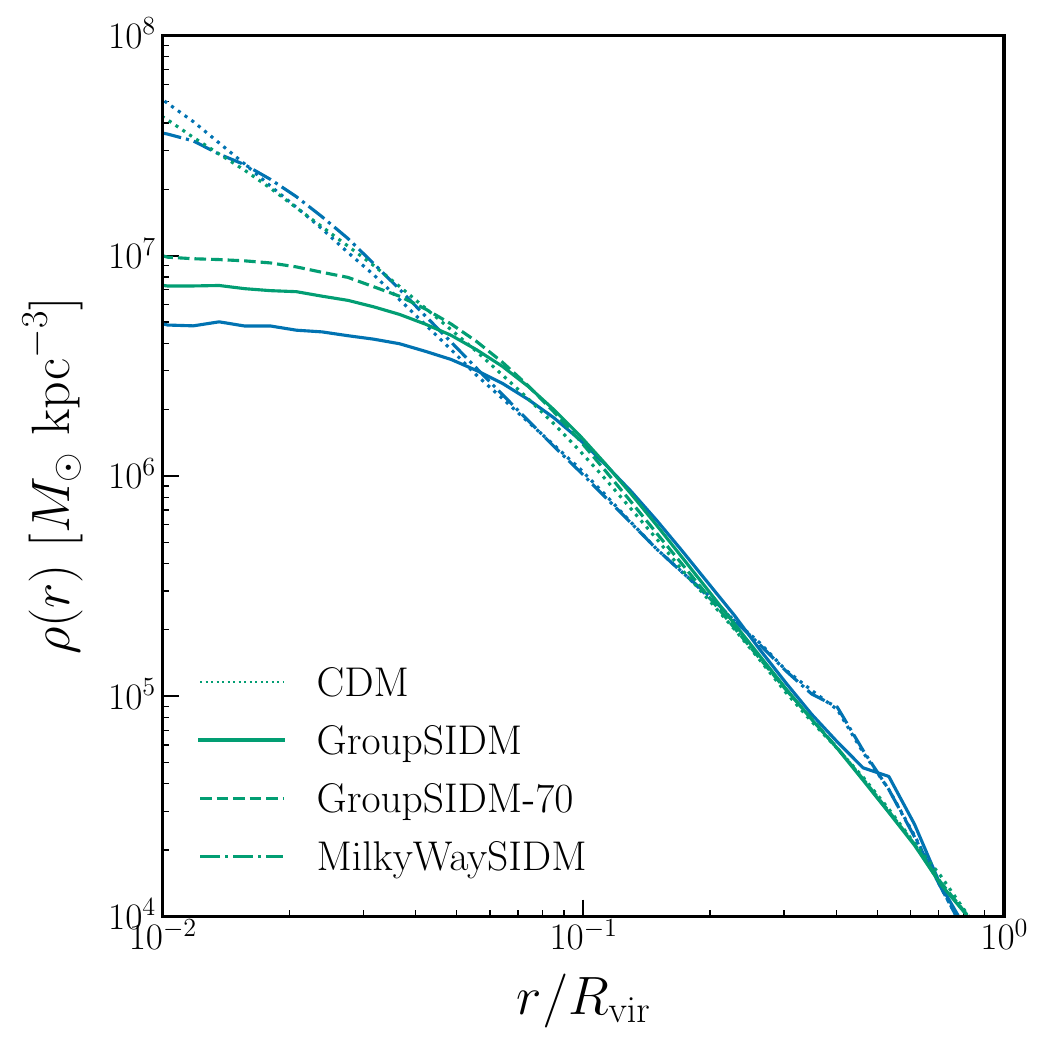}
    \caption{Density profiles of the main host halo in our CDM (dotted) and GroupSIDM (solid) simulations across all suites (left panel) and in the GroupSIDM (solid) and GroupSIDM-70 (dashed) models for the Halo352 Group simulation and the Halo004 MW simulation. Distances are normalized to the virial radius of each host (see Table~\ref{tab:summary} for the $R_{\mathrm{vir}}$ values of each host).}
    \label{fig:host_density}
\end{figure*}

Figure~\ref{fig:host_density} shows the density profile of the main host in each SIDM Concerto simulation. The CDM profiles are roughly self-similar when distances are normalized by the host's virial radius, with the expected trend that lower-mass hosts have more concentrated profiles with higher inner densities. The effect of our GroupSIDM model, shown in the left panel of Figure~\ref{fig:host_density}, is also roughly universal for the lower-mass hosts, which feature density cores within $\approx 0.1R_{\mathrm{vir}}$; the size of the core slightly decreases with decreasing host mass. The L-Cluster host is less affected since it probes a lower-amplitude part of the GroupSIDM cross section (see Figure~\ref{fig:xsec}), but it nonetheless displays a small core within $\approx 3\times10^{-2}R_{\mathrm{vir}}$.

The right panel of Figure~\ref{fig:host_density} compares host density profiles for different SIDM models in Halo004 and Halo352. Decreasing the cross-section amplitude (i.e., changing from GroupSIDM to GroupSIDM-70) or the velocity scale where it transitions to a $v^0$ scaling (i.e., changing from GroupSIDM to MilkyWaySIDM) raises the inner density and decreases the core size. In both cases, these results are expected since the effective cross section at the host's velocity scale is lower in the GroupSIDM-70 and MilkyWaySIDM models than in GroupSIDM (see Figure~\ref{fig:xsec}).

\section{Isolated HMFs}
\label{sec:isolated_hmf}

\begin{figure}[t!]
\centering
\includegraphics[width=0.485\textwidth]{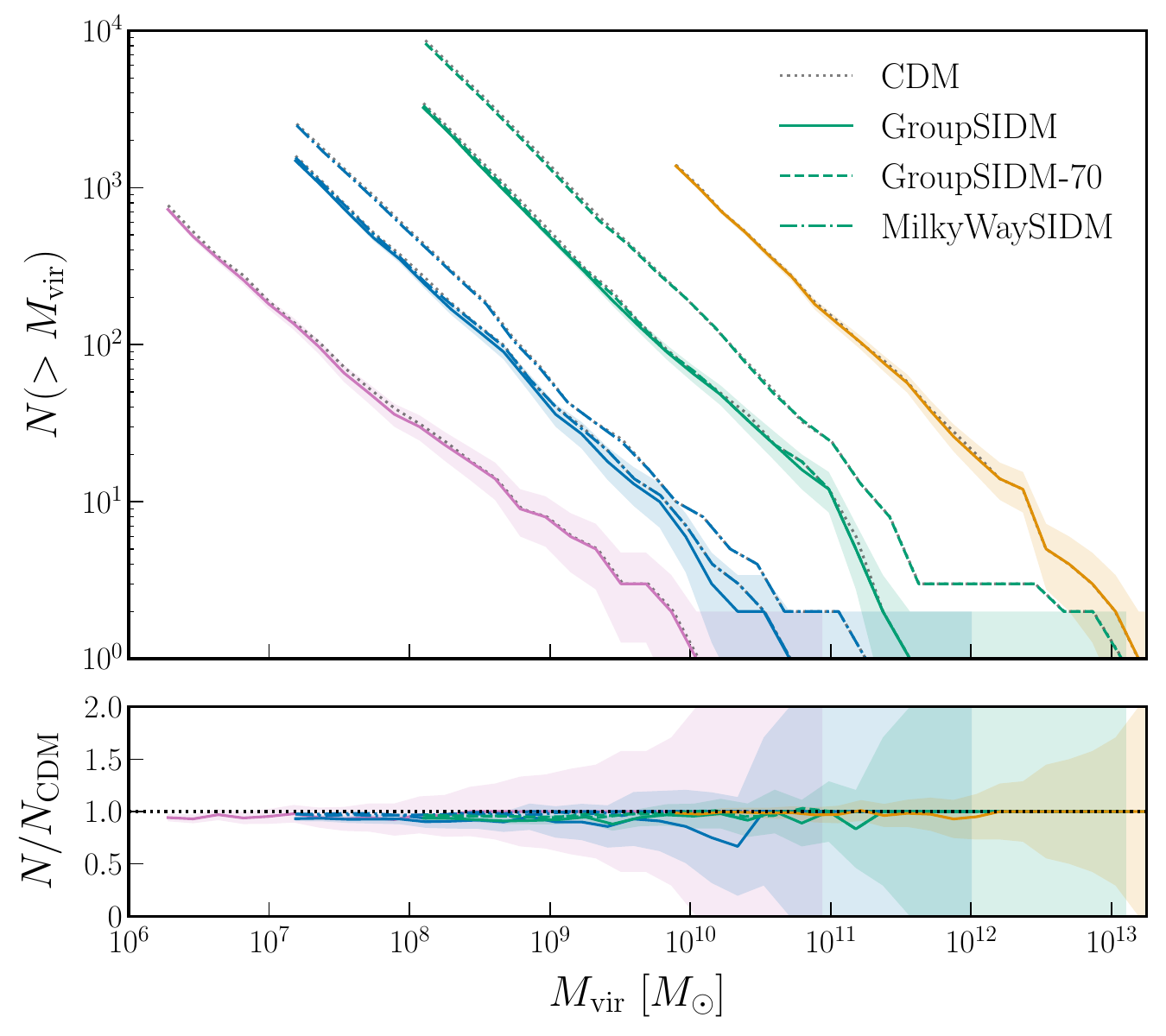}
    \caption{Cumulative isolated HMF calculated using present-day virial mass in SIDM (colored lines) and CDM (dotted black lines). SIDM results are shown for the GroupSIDM (solid), GroupSIDM-70 (dashed), and MilkyWaySIDM (dotted--dashed) models. Shaded bands show the $1\sigma$ Poisson uncertainty on the SIDM SHMFs, and bottom panels show the ratio of the SIDM to CDM SHMF. We restrict to isolated halos within $10~R_{\mathrm{vir}}$ of each host and with $M_{\mathrm{vir}}>300 m_{\mathrm{part}}$.}
    \label{fig:hmf}
\end{figure}

Figure~\ref{fig:hmf} shows the isolated HMFs calculated using \textsc{RCT} for all SIDM Concerto simulations. For these measurements, we select isolated halos that are not within the virial radius of any larger host at $z=0$ (i.e., \texttt{upid}$=-1$). Across all host halo masses and SIDM models, the SIDM HMFs are nearly identical to CDM. This is expected, since the mass accretion histories of isolated SIDM halos are very similar to their CDM counterparts, even though their internal structure (e.g., the $R_{\mathrm{max}}$--$V_{\mathrm{max}}$ relation shown in the right panel of Figure~\ref{fig:rmax_vmax}) can differ from CDM.

\end{document}